\documentclass[12pt]{article}

\usepackage{amsmath,amssymb,amsthm,mathrsfs,bbm}
\usepackage{latexsym,amscd,amsbsy,dsfont,amsfonts}
\usepackage{graphicx}
\usepackage[pdftex,colorlinks=true]{hyperref}

\numberwithin{equation}{section}
\renewcommand{\thefigure}{\arabic{figure}}

\def\eq#1 { \begin{equation} #1 \end{equation} }
\def\eqn#1{ \begin{eqnarray} #1 \end{eqnarray} }
\def\nn { \nonumber }

\def\half{\frac{1}{2}}
\def\Re{{\rm Re}\,}
\def\Im{{\rm Im}\,}

\def\cR{\mathcal{R}}
\def\cL{\mathcal{L}}

\def\d{\partial}

\def\s{\sigma}
\def\a{\alpha}
\def\D{\Delta}
\def\vx{{\vec{x}}}

\def\vL{{\vec{L}}}
\def\vM{{\vec{M}}}
\def\vj{{\vec{j}}}
\def\KG{\overleftrightarrow{\nabla}}

\def\bnu{{\overline{\nu}}}

\def\yb{{\overline{y}}}

\def\C#1{\left\langle #1 \right\rangle}
\def\CP#1{\left\langle #1 \right\rangle_\Psi}
\def\G#1{\Gamma\left(#1\right)}
\def\GG#1{ \Gamma\left[ #1 \right] }
\def \GGG#1#2{\,\Gamma\left[ \begin{array}{l}
      #1 \\
      #2
    \end{array} \right]}
\def\ket#1{\left| #1 \right\rangle}

\def\2F1#1#2#3#4{\,\phantom{}_2F_1\left[#1\,,\,#2\,;\,#3\,;\,#4\right] }
\def\ket#1{\left| #1 \right\rangle}

\begin{document}

\title{The IR stability of de Sitter QFT: Physical initial conditions}

\author{Donald Marolf${}^1$
  \thanks{\href{mailto:marolf@physics.ucsb.edu}{marolf@physics.ucsb.edu}}
    ~ and Ian A. Morrison${}^1$
    \thanks{\href{mailto:ian_morrison@physics.ucsb.edu}{ian\_morrison@physics.ucsb.edu}} \\ \\
    {\it ${}^1$Department of Physics, University of California, Santa Barbara, CA 93016}}

\date{\today}

\maketitle

\begin{abstract}
This work uses Lorentz-signature in-in perturbation theory to analyze the late-time behavior of correlators in time-dependent interacting massive scalar field theory in de Sitter space.  We study a scenario recently considered by Krotov and Polyakov in which the coupling $g$ turns on smoothly at finite time, starting from $g=0$ in the far past where the state is taken to be the (free) Bunch-Davies vacuum.  Our main result is that the resulting correlators (which we compute at the one-loop level) approach those of the interacting Hartle-Hawking state at late times.  We argue that similar results should hold for other physically-motivated choices of initial conditions.  This behavior is to be expected from recent quantum ``no hair'' theorems for interacting massive scalar field theory in de Sitter space which established similar results to all orders in perturbation theory for a dense set of states in the Hilbert space. Our current work i) indicates that physically motivated initial conditions lie in this dense set, ii) provides a Lorentz-signature counter-part to the Euclidean techniques used to prove such theorems, and iii) provides an explicit example of the relevant renormalization techniques.
\end{abstract}

\pagebreak

\tableofcontents
\sloppy

\section{Introduction}

Quantum field theories (QFTs)
in de Sitter space are of interest for many reasons. Some of these include the increasingly precise
measurements of the cosmic microwave background (CMB) \cite{Komatsu:2008hk}
which have prompted many to study predictions of the CMB spectrum
beyond the Born approximation (see, e.g.
\cite{Weinberg:2005vy,Weinberg:2006ac,Seery:2005wm,Seery:2007we,
Cheung:2007st,Senatore:2009cf,Seery:2010kh,Giddings:2010nc,Giddings:2011zd}).
Others include a growing interest in understanding
local measurements in eternal inflation (e.g., \cite{Bousso:2006aa,
  Guth:2007aa,Hartle:2010dq}), as well as
a renewed interest in approaches to de Sitter quantum
gravity \cite{Seery:2006tq,Anninos:2009yc,Anninos:2010gh,Anninos:2011vd} inspired by dS/CFT \cite{Witten:2001kn,Strominger:2001pn}.  In addition, the fact 
that de Sitter (dS) is a maximally symmetric (and thus relatively simple) 
example of a spacetime where horizons limit the observations of freely-falling 
observers makes QFTs on dS of interest in their own right.

One of the chief concerns with QFTs in de Sitter has been their infrared stability
(see e.g. \cite{
Nachtmann:1968aa,Myhrvold:1983hx,Hu:1985uy,Hu:1986cv,Boyanovsky:1996ab,Bros:2006gs,
Polyakov:2007mm,Akhmedov:2008pu,Higuchi:2008tn,Higuchi:2009zza,Higuchi:2009ew,Akhmedov:2009ta,Polyakov:2009nq,
Burgess:2010dd,Marolf:2010zp,Rajaraman:2010zx,Youssef:2010dw,Boyanovsky:2011xn,Hollands:2010pr,
  Marolf:2010nz,Krotov:2010ma}).
Two recent papers \cite{Hollands:2010pr,Marolf:2010nz} have established that at least massive scalar field quantum field theories are infra-red stable (at all orders of perturbation theory) in a particular sense. In order to state their results precisely, we first introduce the (interacting) Hartle-Hawking state $\ket{\rm HH}$
\cite{Hartle:1976tp} defined by analytically continuing all correlation functions from Euclidean signature.  Next, consider a normalized state constructed by the application of smeared field operators on $\ket{\rm HH}$:
\eq{ \label{eq:genericPsi}
  \ket{\Psi} = \int_{y_1}\cdots\int_{y_n} f(y_1,\dots,y_n)
  \phi(y_1)\cdots\phi(y_n) \ket{\rm HH} ,
}
with $f(y_1,\dots,y_n)$ a smooth smearing function of compact
support. By the Reeh-Schlieder theorem of curved spacetimes
\cite{strohmaier:5514} the set of states of the form (\ref{eq:genericPsi})
is dense in the Hilbert space. The works \cite{Hollands:2010pr,Marolf:2010nz}
show that, at all orders of perturbation theory, the correlation functions of $\ket{\Psi}$ reduce to
those of the Hartle-Hawking state when evaluated in the asymptotic
future/past of de Sitter space:
\eq{\label{eq:nh}
  \C{\phi(x_1)\cdots\phi(x_n)}_\Psi \to
  \C{\phi(x_1)\cdots\phi(x_n)}_{\rm HH} .
}
In particular, de Sitter invariance of $|{\rm HH} \rangle $ means that all one-point functions approach constants (whether the associated operators are elementary or composite). Ref \cite{Hollands:2010pr} calls the result (\ref{eq:nh})
a `quantum cosmic no hair theorem,' while in the language of \cite{Marolf:2010zp} one says that $|{\rm HH} \rangle$ is an attractor state for local correlators.

Although stated as a result concerning QFTs in exact de Sitter
space, the no-hair theorem just described
may also be usefully applied to more interesting scenarios. Since it relies only on the asymptotic behavior, it should be valid in the asymptotic region of
any asymptotically-de Sitter spacetime, or within the causal patch
of an observer who finds herself in a locally de Sitter spacetime.
One expects physically relevant states to take the form (\ref{eq:genericPsi}) within the de Sitter region so that the theorem applies.

Our main purpose here is to provide evidence that this is indeed the case by studying (at the one-loop level) a particular scenario recently discussed by Krotov and Polyakov \cite{Krotov:2010ma}.   In this scenario, the spacetime is again exact de Sitter but the theory is time dependent.  The particular model involves a cubic interaction
$g(x) \phi^3(x)$ with time-dependent coupling $g(x)$, taken to vanish in the far past and to approach some constant $g_f$ in the far future.  The state of the system is taken to be the free Bunch-Davies vacuum in the region where $g(x) =0$ and we take the coupling to turn on at some fixed time.  We explicitly compute the $O(g^2)$ (one-loop) corrections to the 2-point function
in this model and verify that they approach those of the
de Sitter Hartle-Hawking state in the far future.  We work entirely in Lorentz signature, in part to counter concerns \cite{Polyakov:2007mm,Polyakov:2009nq,Krotov:2010ma} about the Euclidean techniques used in \cite{Hollands:2010pr,Marolf:2010nz}.  In addition, we note that the current work provides an an explicit example of the renormalization techniques used in \cite{Marolf:2010nz} which combine Pauli-Villars regularization with Mellin-Barnes representations.  Our techniques also apply to other examples where the coupling does not depend on time but in which the spacetime is de Sitter only after some finite time.

Although \cite{Krotov:2010ma} concluded that de Sitter QFT is ``unstable,'' it is useful to point out that our technical results are completely consistent with those of \cite{Krotov:2010ma}.  As stated below their equation (17), for fixed $g(x)$ their approximations are not valid for correlators computed at late times.  Instead, \cite{Krotov:2010ma} focused on correlators defined at some {\it fixed} time in the limit where $g(x)$ turns on at very early times.  The divergence they find is in fact to be  expected from the results of \cite{Hollands:2010pr,Marolf:2010nz}, which suggest that correlators in well-behaved  states approach those of $|{\rm HH} \rangle$ in the far past.  Using the free vacuum when $g(x)=0$ and taking $g(x)$ to turn on very early ensures that correlators at such early times differ significantly from those in $|{\rm HH} \rangle$.  As a result, one already expects from \cite{Hollands:2010pr,Marolf:2010nz} that the state is not well-behaved under the limit taken in \cite{Krotov:2010ma}.

We emphasize that the present work considers only perturbative effects in massive theories.  Non-perturbative effects can yield qualitatively different behavior (though see \cite{Shlaer:2009aa}), and
including massless fields (whether scalar or tensor\footnote{The 2-point 
functions of Maxwell fields in dS are known to behave like those of 
massive scalars 
\cite{Allen:1985wd,Tsamis:2006gj} and so provide no new subtleties.}) 
would raise new issues. As a result, our analysis does not directly address the much-discussed possibility of novel infrared effects in de Sitter quantum gravity
(e.g., \cite{Tsamis:1992sx,Tsamis:1994ca,Mottola:2006ew,Antoniadis:2006wq,
  Garriga:2007zk,Tsamis:2007is,Urakawa:2010it,Tsamis:2011uq,Giddings:2010nc,Giddings:2011zd}).
Nevertheless, a detailed study of theories on a fixed de Sitter background helps to disambiguate those effects which are truly quantum gravitational from
those that are generic for quantum fields in de Sitter.

We begin in \S\ref{sec:prelims}
with a brief review of linear quantum field theory on de Sitter.  We then analyze
time-dependant couplings in global de Sitter in \S\ref{sec:time}. Section 
\ref{sec:phi2} studies a simple theory with a time-dependant $\phi^2(x)$ interaction (i.e., a time-dependant mass perturbation). We study the more complicated model of a time-dependant $\phi^3(x)$ interaction in \S\ref{sec:phi3}, relying
at times upon the results derived in the $\phi^2(x)$ model.
Some further calculational details are presented in appendices.
We provide a concluding discussion in \S\ref{sec:disc}.

\section{Preliminaries}
\label{sec:prelims}

We begin by reviewing some basic features of {\it free} quantum fields in de Sitter space.  Recall that global de Sitter
may be described by the metric
\eq{ \label{eq:metric}
  ds^2 = \ell^2\left[ - \frac{1}{1+\eta^2} d\eta^2
    + (1+\eta^2) d\Omega_{D-1}^2 \right] ,
}
where $\ell$ is the de Sitter radius, $\eta$ is a time coordinate
with range $-\infty < \eta < + \infty$, and $d\Omega^2_{D-1}$ is the
metric of the unit sphere $S^{D-1}$. The coordinate $\eta$ is related
to the more familiar global de Sitter coordinate $t$ (for which $g_{tt}=-1$) via
$\eta = \sinh(t/\ell)$. In these coordinates the volume element is
$\sqrt{-g(x)} d^Dx = \ell^D (1+\eta^2)^{(D-2)/2} d\eta \,d\Omega_{D-1}(\vx)$
with $\vx$ a unit vector in $\mathbb{R}^{D-1}$ parameterizing $S^{D-1}$.
The Euclidean section of de Sitter is the Euclidean sphere $S^D$ with
radius $\ell$.

Free massive scalar field theories on de Sitter have been well-understood
for decades (e.g., \cite{Allen:1985ux,Birrell:1982ix}). Such
theories may be described by the classical Lagrangian density
\eq{
  \cL = \half \nabla_\mu \phi \nabla^\mu \phi + \frac{M^2}{2}\phi^2 ,
}
from which it follows that the classical equation of motion is the
Klein-Gordon equation; correspondingly, in the quantum theory the
Schwinger-Dyson equations are
\eq{ \label{eq:FreeSDEquations1}
  (\Box_i - M^2) \CP{\phi(x_1)\cdots\phi(x_i)\cdots\phi(x_N)} = 0 .
}
The free theory admits a unique Hadamard de Sitter-invariant state known
as the Bunch-Davies, Euclidean, or (free) Hartle-Hawking state, which we
denote by $\ket{0}$.
The latter two names come from the fact that the correlation functions
of this state may be defined by the analytic continuation from the Euclidean
section.
We denote the time-ordered, anti-time-ordered, and Wightman 2-point
functions of this state by
\eqn{ \label{eq:Greens}
  G_\s(x_1,x_2) &:=& \C{ T \phi_\s(x_1)\phi_\s(x_2) }_0 , \nn \\
  G^*_\s(x_1,x_2) &:=&
  \C{ \overline{T} \phi_\s(x_1)\phi_\s(x_2) }_0 , \nn \\
  W_\s(x_1,x_2) &:=& \C{ \phi_\s(x_1)\phi_\s(x_2)}_0 .
}
In these expressions we have introduced a label $\s$ to keep track of
the bare mass $M$. We define $\s$ by
\eq{
  \s = - \left(\frac{D-1}{2}\right)
  + \left[\frac{(D-1)^2}{4} - M^2\ell^2 \right]^{1/2} ,
}
from which it follows that $M^2 \ell^2 = - \s(\s + D-1)$.

At times it will be convenient to expand the scalar Green's functions
(\ref{eq:Greens}) in Klein-Gordon modes $\phi_{\s \vL}(x)$. These
modes are orthonormal with respect to the Klein-Gordon inner product:
\eq{ \label{eq:KGnorm}
  \left( \phi_{\s \vL},\,\phi_{\s \vM} \right)_{\rm KG} :=
  - i \ell^{D-1} (1+\eta^2)^{(D-1)/2} \int d\Omega_{D-1}(\vx) \,n^\mu
  \left[ \phi_{\s \vL}(\eta,\vx) \KG_\mu \phi^*_{\s \vM}(\eta,\vx) \right]
  = \delta_{\vL \vM} .
}
Here $n^\mu$ is the future-directed normal vector
($n^\mu = (1+\eta^2)^{1/2} \delta^\mu_\eta/\ell$) to an $\eta={\rm const.}$ surface
and $A \KG_\mu B := A \nabla_\mu B - B \nabla_\mu A$.
The Klein-Gordon modes may be written explicitly as
\eq{ \label{eq:KGmodes}
  \phi_{\s \vL}(x) = \ell^{(2-D)/2} u_{\s L}(\eta) Y_{\vL}(\vx) ,
}
with $Y_\vL(\vx)$ spherical harmonics on $S^{D-1}$ and $u_{\s L}(\eta)$
given by
\eq{ \label{eq:u}
  u_{\s L}(\eta) = N_{\s L} (1+\eta^2)^{-(D-2)/4}
  \left[\frac{1-i\eta}{1+i\eta}\right]^{(L+(D-2)/2)/2}
  F_{\s L}(\eta) ,
}
where $F_{\s L}(\eta)$ is a Gauss hypergeometric function
\eq{ \label{eq:F}
  F_{\s L}(\eta) =
  \2F1{\s+\frac{D}{2}}{1 - \s -\frac{D}{2}}{L+\frac{D}{2}}{\frac{1-i\eta}{2}} ,
}
and the normalization coefficient is
\eq{ \label{eq:N}
  N_{\s L} = \frac{1}{\G{L+\frac{D}{2}}}
  \left[
    \frac{\GG{L-\s,L+\s+D-1}}{2}
  \right]^{1/2} .
}

Using these modes we may expand, e.g., the Wightman function
\eqn{ \label{eq:Wmodes}
  W_\s(x_1, x_2) &=&
  \ell^{2-D} \sum_\vL \phi_{\s \vL}(x_1) \phi^*_{\s \vL}(x_2)
  \nn \\
  &=& \ell^{2-D} \frac{\G{\frac{D-2}{2}}}{4 \pi^{D/2}}
  \sum_{L=0}^\infty (2L+D-2)
  u_{\s L}(\eta_1) u^*_{\s L}(\eta_2) C_L^{(D-2)/2}(\vx_1\cdot\vx_2) .
}
Here $C_L^{(D-2)/2}(z)$ is a Gegenbauer polynomial and once again
$\vx_1,\vx_2$ are unit vectors in $\mathbb{R}^{D-1}$ parameterizing the
$S^{D-1}$. To obtain the last equality we sum over angular momenta via
the useful identity
\eq{
  \sum_\vj Y_\vL(\vx_1) Y^*_\vL(\vx_2) =
  \frac{\G{\frac{D-2}{2}}}{4 \pi^{D/2}}
  (2L+D-2) C_L^{(D-2)/2}(\vx_1\cdot\vx_2), \quad \vL = (L,\vj) .
}

It will be useful to note some qualitative features of the
of the Klein-Gordon mode function $u_{\s L}(\eta)$. First, as one might expect from the fact that the volume of the $S^{D-1}$ is smallest at $\eta =0$, the mode  functions are
bounded by their values at $\eta = 0$:
\eqn{ \label{eq:ubound}
  |u_{\s L}(\eta)|^2 \le |u_{\s L}(0)|^2
  &=& 2^{-(2L+D-1)}\pi \frac{\GG{L-\s, L+\s+D-1}}
  {\left(\GG{\frac{1+L-\s}{2}, \frac{1+L+\s+D-1}{2}}\right)^2}
  \nn \\
  &=& \frac{1}{2L}\left(1 + O(L^{-1})\right) ,
  \quad {\rm when\;} L \gg 1,
  \;\; \s\; {\rm fixed \;} ,
}
(see eq. (56) of \cite{Marolf:2008hg}).
From this we see that $|u_{\s L}(\eta)|$ may be bounded by a function of
$L$ that decreases as $L \to \infty$.

Second, the expansion of the universe and the ensuing growth of the physical wavelength at fixed $L$ suggest that at large $|\eta|$ all modes behave like the $L=0$ mode.  Indeed, as derived in detail in Appendix \ref{app:asymptotic}, the following asymptotic expansion for $u_{\s L}(\eta)$ is
valid for large $|\eta|$ when $\s$, $L$, and $D$ are held fixed:
\eqn{ \label{eq:uLargeEta}
  u_{\s L}(\eta) &=& \frac{N_{\s L}}{2^{\s+(D-2)/2}}
  \GGG{L+\frac{D}{2}, 2\s+D-1}{L+\s+D-1, \s+\frac{D}{2}}
  \exp\left[i\frac{\pi}{2}(L+\s+D-2)\right] (\eta)^{\s}
  \left[1 + O\left(\frac{(L-\s)}{\eta}\right)\right]
  \nn \\ & &
  + (\s \to - (\s+D-1)) ,
  \quad {\rm for\;} |\eta| \gg 1, \;\; |\eta| \gg (L-\s) .
}

Now, while (\ref{eq:uLargeEta}) gives the correct asymptotic behavior for
a given mode at large $|\eta|$ (with all other parameters fixed), it
does not correctly reproduce the behavior of the mode function at some
arbitrarily large-but-finite $|\eta|$ as $L \to \infty$. To understand
the behavior of the mode function in this regime, we instead use the
WKB approximation valid when
\eq{
  f(\eta) := \frac{L(L+D-2)}{(1+\eta^2)} + M^2\ell^2 \gg 1 .
}
The WKB approximation is derived in appendix~\ref{app:WKB} and is
given by
\eq{ \label{eq:WKB}
  u_{\s L}(\eta) \approx \frac{1}{\sqrt{2}}
  (1+\eta^2)^{(1-D)/4}
  \left[ f(\eta) \right]^{-1/4} e^{\pm i \Upsilon(\eta) } ,
}
with $\Upsilon(\eta)$ satisfying
\eq{
  \frac{d}{d\eta} \Upsilon(\eta)
  = \left[\frac{f(\eta)}{(1+\eta^2)}\right]^{1/2} .
}
The key feature of this expression is that $\Upsilon(\eta)$ is
large in the regime of validity, so (\ref{eq:WKB}) is a highly oscillatory
function of $\eta$.

Let us now turn the discussion to interacting theories.  The Hartle-Hawking state $\ket{\rm HH}$ is constructed by analytically continuing all correlation functions from the Euclidean section. We denote the Hartle-Hawking
state constructed perturbatively in an interacting theory by
$\ket{\rm HH}$, and reserve $\ket{0}$ to denote the Hartle-Hawking
state of the free theory. The state $\ket{\rm HH}$  has been studied in detail for
massive scalar field theories in \cite{Higuchi:2009ew,Marolf:2010zp,Hollands:2010pr,Marolf:2010nz,
  Higuchi:2010aa} by performing the relevant analytic continuations. However, in the current work we perform all
calculations explicitly in Lorentz-signature using standard
Schwinger-Keldysh (a.k.a ``in-in'', ``real-time'', ``closed time path'')
perturbation theory
(for original works see \cite{Schwinger:1960qe,Keldysh:1964ud};
for more tractable introductions see
\cite{Jordan:1986ug,Paz:1990jg,Vilkovisky:2007ny} and the appendix
of \cite{Weinberg:2005vy}).

\section{Time-dependent couplings in de Sitter}
\label{sec:time}

\begin{figure}
  \centering
  \includegraphics{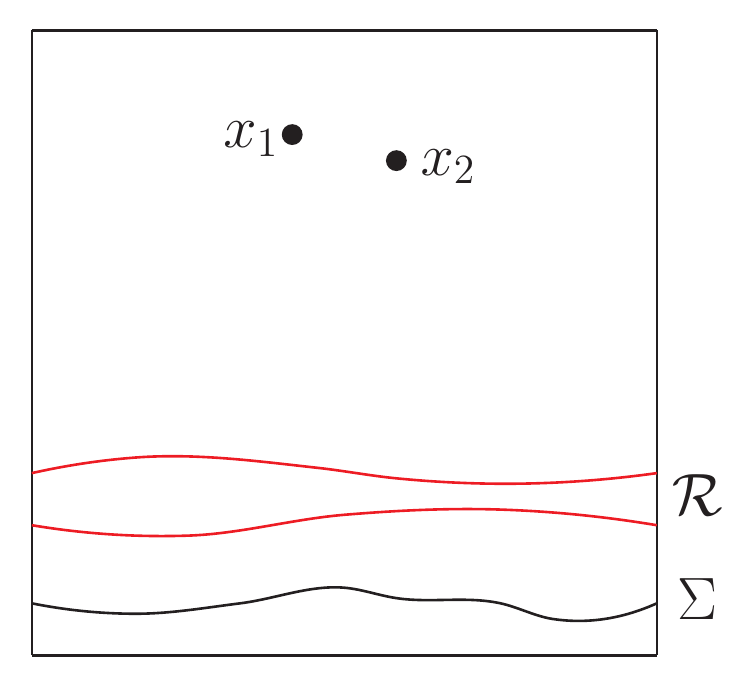}
  \caption{The Penrose diagram of de Sitter. We consider the
    state $\ket{\Psi}$ defined by the Bunch-Davies vacuum on a Cauchy
    surface $\Sigma$. The interaction turns on across the region
    $\mathcal{R}$ via a smooth coupling function $g(x)$ such that $g(x)=g_f$
    in the causal future of $\mathcal{R}$ and $g(x)=0$ in the
    causal past of $\mathcal{R}$. We compute the time-ordered 2-point
    function $\CP{T\phi(x_1)\phi(x_2)}$ of two points in the distant future.
  }
  \label{fig:dS}
\end{figure}
\renewcommand{\thefigure}{\arabic{figure}}

Consider a massive scalar field
on global de Sitter with a time-dependent self-interaction.
In particular, let the self-interaction vanish in the asymptotic past but turn on smoothly across a spacetime region $\cR$.  We require the coupling function $g(x)$ to satisfy
\eq{
  g(x) = \left\{ \begin{array}{ll}
      0 & \quad {\rm for\;} x \in J^-(\cR)/\cR  \\
      g_f & \quad {\rm for\;} x \in J^+(\cR)/\cR
      \end{array} \right. ,
}
where as usual $J^\pm$ denotes the causal past and future of a set and $A/B$ denotes the set of points in $A$ that do not lie in $B$.  We sketch the scenario in Fig.~\ref{fig:dS}.

We wish to compute correlation functions with respect to the state
$\ket{\Psi}$ which coincides with the free Bunch-Davies vacuum
$\ket{0}$ in the past region $J^+(\cR)/\cR$.
For infinitesimal coupling $g(x)$ the time-ordered 2-point function
with respect to $\ket{\Psi}$ can be expanded as
\eq{ \label{eq:2pt}
  \CP{ T \phi(x_1)\phi(x_2) }
  = \C{ T \phi(x_1)\phi(x_2) }_0
  + \sum_{n=1}^\infty \CP{T \phi(x_1)\phi(x_2)}^{(n)} ,
}
where $\CP{T \phi(x_1)\phi(x_2) }^{(n)}$ is of $O(g^n)$.
We choose a surface $\Sigma$ in the past of $\cR$ as our initial
Cauchy surface -- see Fig.~\ref{fig:dS}. Given the above
choice of state, the appropriate Green's functions to use in the
Schwinger-Keldysh formalism are those of the
Bunch-Davies vacuum (\ref{eq:Greens}).

\begin{figure}
  \centering
  \includegraphics{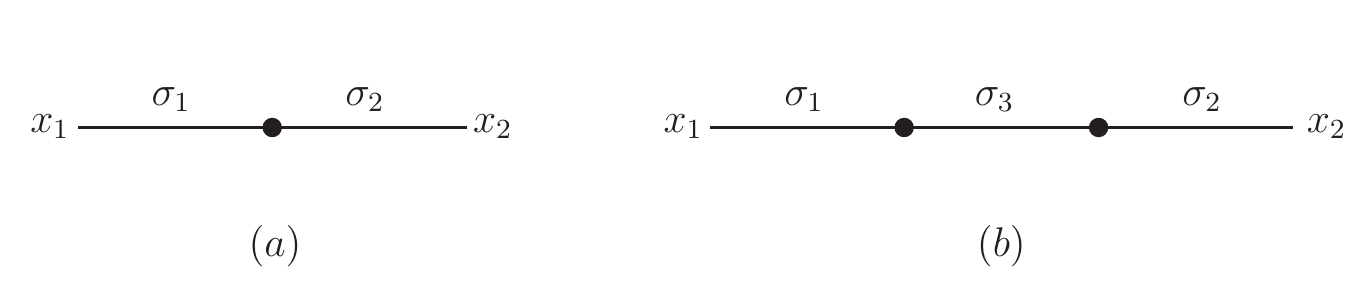}
  \caption{Corrections to the time-ordered 2-point function
    $\CP{T \phi(x_1)\phi(x_2)}$ due to the interaction $-\frac{1}{2}g(x)\phi^2(x)$. The $O(g(x))$ correction is depicted in Fig. (a)
    while the $O(g^2(x))$ correction is depicted in Fig. (b). It is
    convenient for computation to label each leg of the diagram by a
    distinct mass parameter $\s_i$.}
  \label{fig:phi2}
\end{figure}

The main result of this section is to show that, when it is
evaluated at $x_1$, $x_2$ in the far future of $\cR$, the 2-point function
(\ref{eq:2pt}) reduces to that of the Hartle-Hawking state of the
analogous theory with constant coupling $g_f$. We
consider both quadratic and cubic couplings ($g(x)\phi^2(x)$ and $g(x) \phi^3(x)$) below, showing in each case that the perturbative corrections
$\C{T \phi(x_1)\phi(x_2)}_\Psi^{(n)}$ approach those of the Hartle-Hawking state at late times up to order $n=2$.  Although it may be of less physical interest, the analysis of the quadratic coupling in section \ref{fig:phi2} will help to greatly simplify our 1-loop treatment of the cubic coupling in section  \ref{fig:phi3}.

\subsection{Example: $\phi^2(x)$ interaction}
\label{sec:phi2}

\renewcommand{\thefigure}{\arabic{figure}}

In our first example we consider the simple quadratic interaction term
\eq{
  \cL_{\rm int}[\phi] = - \frac{g(x)}{2} \phi^2(x) .
\label{eq:Lp2}}
Treating this term perturbatively yields only tree-level diagrams; no
regularization
or renormalization is needed.   We compute both the $O(g)$ and the $O(g^2)$ corrections to the two-point function of this model below.  As we will see in section \ref{sec:phi3}, the $O(g^2)$ correction from (\ref{eq:Lp2}) is closely related to the 1-loop correction from a $\phi^3$ interaction.  As a result, the results below will greatly simplify the manipulations in section \ref{sec:phi3}.

\subsubsection{The $O(g)$ correction}
\label{sec:Og}

The $O(g)$ correction is depicted in the
Feynman diagram  shown in Fig.~\ref{fig:phi2}~(a) and is given by the expression
\eq{ \label{eq:phi2}
  \CP{T \phi_{\s_1}(x_1) \phi_{\s_2}(x_2)}^{(1)}
  = i \int_y g(y)\left[ G_{\s_1}(y,x_1) G_{\s_2}(y,x_2)
    - W_{\s_1}(y,x_1) W_{\s_2}(y,x_2)\right] .
}
We denote by $\int_y\dots$ an integral over the future of $\Sigma$.
For the moment it is convenient to let each Green's function have a
distinct mass; we will take the limit of equal masses later.

Consider the first term in (\ref{eq:phi2}):
\eq{ \label{eq:T1}
  T_1(x_1,x_2) := i \int_y g(y) G_{\s_1}(y,x_1) G_{\s_2}(y,x_2) .
}
Making use of the Green's functions' equations of motion
\eqn{
  (\Box_x - M^2) G_\s(x,y) &=& (\Box_y - M^2) G_\s(x,y) = i \delta(x,y),
  \nn \\
  (\Box_x - M^2) W_\s(x,y) &=& (\Box_y - M^2) W_\s(x,y) = 0 ,
}
we may usefully re-write (\ref{eq:T1}) as
\eqn{
  T_1(x_1,x_2) &=&
  \frac{i}{M_1^2 - M_2^2} \int_y g(y) \bigg[
    (\Box_y G_{\s_1}(y,x_1))G_{\s_2}(y,x_2)
    \nn \\ & &
    - G_{\s_1}(y,x_1)(\Box_y G_{\s_2}(y,x_2))
    - i \delta(x_1,y) G_{\s_2}(y,x_2)
    + i \delta(x_2,y) G_{\s_1}(y,x_1) \bigg]
    \nn \\ &=&
    \frac{1}{M_1^2-M_2^2}\left[ g(x_1) G_{\s_2}(x_1,x_2)
      - g(x_2) G_{\s_1}(x_1,x_2) \right]
    \nn \\ & &
    + \frac{i}{M_1^2 - M_2^2} \int_y g(y) \bigg[
    (\Box_y G_{\s_1}(y,x_1))G_{\s_2}(y,x_2)
    - G_{\s_1}(y,x_1)(\Box_y G_{\s_2}(y,x_2)) \bigg] .
    \nn \\
}
Making the simple re-arrangement
\eqn{
  g(y) (\Box G_{\s_1}(y,x_1)) G_{\s_2}(y,x_2)
  &=&
  \nabla^\mu \big[ g(y) (\nabla_\mu G_{\s_1}(y,x_1)) G_{\s_2}(y,x_2)\big]
  \nn \\ & &
  - (\nabla^\mu g(y)) (\nabla_\mu G_{\s_1}(y,x_1)) G_{\s_2}(y,x_2)
  \nn \\ & &
  - g(y) (\nabla_\mu G_{\s_1}(y,x_1)) (\nabla^\mu G_{\s_2}(y,x_2)) ,
}
we obtain
\eqn{ \label{eq:T1b}
  T_1(x_1,x_2) &=&
    \frac{1}{M_1^2-M_2^2}\left[ g(x_1)G_{\s_2}(x_1,x_2)
      - g(x_2)G_{\s_1}(x_1,x_2) \right]
    \nn \\ & &
    - \frac{i}{M_1^2-M_2^2} \int_y \nabla^\mu
    \left[ g(y) G_{\s_1}(y,x_1) \KG_\mu G_{\s_2}(y,x_2) \right]
    \nn \\ & &
    + \frac{i}{M_1^2-M_2^2} \int_y (\nabla^\mu g(y) )
    \left[ G_{\s_1}(y,x_1) \KG_\mu G_{\s_2}(y,x_2) \right] .
}
One can perform the same
manipulations for the second term in (\ref{eq:phi2}). The only difference
is that the Wightman function satisfies the homogeneous equation of motion,
so there are no analogs of the terms on the top line of (\ref{eq:T1b}).
All together we obtain the expression
\eqn{ \label{eq:phi2b}
  & &\CP{T \phi_{\s_1}(x_1) \phi_{\s_2}(x_2)}^{(1)}
  =
  \frac{1}{M_1^2-M_2^2}\left[ g(x_1)G_{\s_2}(x_1,x_2)
    - g(x_2) G_{\s_1}(x_1,x_2) \right]
  \nn \\ & &
  - \frac{i}{M_1^2-M_2^2} \int_y \nabla^\mu
  \left[ g(y) \left( G_{\s_1}(y,x_1) \KG_\mu G_{\s_2}(y,x_2)
      - W_{\s_1}(y,x_1) \KG_\mu W_{\s_2}(y,x_2) \right) \right]
  \nn \\ & &
  + \frac{i}{M_1^2-M_2^2} \int_y (\nabla^\mu g(y) )
  \left[ G_{\s_1}(y,x_1) \KG_\mu G_{\s_2}(y,x_2)
    - W_{\s_1}(y,x_1) \KG_\mu W_{\s_2}(y,x_2)
  \right] .
}

In the second line of (\ref{eq:phi2b}) there is an integral of a total
derivative. By Stokes' theorem this integral can be expressed as an
integral over the boundary of the region to the future of $\Sigma$. This boundary
is simply the union of $\Sigma$ and future infinity $I^+$.
Now, the combination of Green's functions in the integrand is such that
the integrand has support only on the union of the past light cones of
$x_1$ and $x_2$, so the integral over $I^+$ vanishes. Furthermore, the
coupling function $g(y)$ vanishes on $\Sigma$, so the integral over
$\Sigma$ vanishes as well. We conclude that the integral in the
second line of (\ref{eq:phi2b}) is identically zero:
\eqn{ \label{eq:phi2c}
  & &\CP{T \phi_{\s_1}(x_1) \phi_{\s_2}(x_2)}^{(1)}
  =
  \frac{1}{M_1^2-M_2^2}\left[ g(x_1) G_{\s_2}(x_1,x_2)
    - g(x_2) G_{\s_1}(x_1,x_2) \right]
  \nn \\ & &
  + \frac{i}{M_1^2-M_2^2} \int_y (\nabla^\mu g(y) )
  \left[ G_{\s_1}(y,x_1) \KG_\mu G_{\s_2}(y,x_2)
    - W_{\s_1}(y,x_1) \KG_\mu W_{\s_2}(y,x_2)
  \right] .
}

We are interested in $x_1,x_2$ in the future region $J^+(\cR)/\cR$.  Since the gradient $\nabla^\mu g$ has support only in $\cR$,  we need not allow $y$ in (\ref{eq:phi2c}) to coincide with $x_1,x_2$ or to lie in the future of either point.  We may therefore replace $G_{\sigma_1}, G_{\sigma_2}$ by appropriate Wightman functions in the integral and write the second line of (\ref{eq:phi2c}) in the form
\eqn{ \label{eq:T2}
  T_{2\,\s_1\s_2}(x_1,x_2) &:=&
\frac{i}{M_1^2-M_2^2} \int_y (\nabla^\mu g(y) )
  \left[ W_{\s_1}(x_1,y) \KG_\mu W_{\s_2}(x_2,y)
    - W_{\s_1}(y,x_1) \KG_\mu W_{\s_2}(y,x_2)
  \right]
  \nn \\
  &=& \frac{2}{M_1^2 - M_2^2}
  \Im\left\{
      \int_y (\nabla^\mu g(y) ) W_{\s_1}(y,x_1) \KG_\mu W_{\s_2}(y,x_2)
    \right\} .
}
To keep the computation simple we choose $g(y)$ to be a smooth
function of the time coordinate $\eta$ alone, i.e.,
\eq{ \label{eq:geta}
  \nabla^\mu g(y) = - \ell^{-2} (1+\eta^2) g'(\eta) \delta^\mu_\eta .
}
By expanding the Wightman functions in Klein-Gordon modes as in
(\ref{eq:Wmodes}) we compute
\eqn{
  T_{2\,\s_1\s_2}(x_1,x_2)
  &=& - \frac{2 \ell^{D-2}}{M_1^2 - M_2^2}
  \Im\left\{
    \int d\eta \, (1+\eta^2)^{(D-1)/2} g'(\eta) \int d\Omega_{D-1}
    W_{\s_1}(y,x_1) \overleftrightarrow{\d_n} W_{\s_2}(y,x_2)
    \right\}
    \nn \\
  &=& - \ell^{2-D} \frac{\G{\frac{D-2}{2}}}{2\pi^{D/2}}
      \Im
      \sum_{L=0}^\infty \bigg\{ (2 L+D-2) \chi_{\s_1\s_2}(L)\,
      u^*_{\s_1 L}(t_1) u^*_{\s_2 L}(t_2)
      C^{(D-2)/2}_L(\vx_1\cdot\vx_2)
      \bigg\}  , \nn \\
      \label{eq:T2again}
}
with
\eq{ \label{eq:chi}
  \chi_{\s_1\s_2}(L)
  = \frac{1}{(M_1^2-M_2^2)} \int d\eta\, (1+\eta^2)^{(D-1)/2}
  g'(\eta) \left[
    u_{\s_1 L}(\eta) \overleftrightarrow{\d_n}
    u_{\s_2 L}(\eta) \right] ,
}
where $\d_n$ denotes a derivative along the unit (future-pointing)
timelike normal to the surface $\eta = {\rm const}$.
The integral in (\ref{eq:chi}) is guaranteed to be finite for any
pair $u_{\s_1 L}(\eta)$, $u_{\s_2 L}(\eta)$ as the harmonics are bounded
and $g'(\eta)$ is smooth with compact support.
It is also finite in the limit $M_2^2 \to M_1^2$, as can be seen by
using l'Hopital's rule.

It remains, however, to determine the convergence of the sum over $L$
in (\ref{eq:T2again}). This is governed by the behavior of $\chi_{\s_1\s_2}(L)$
at large $L \gg 1$. Since $g'(\eta)$ has compact support it must vanish for
$|\eta|$ greater than some $|\eta|_{\rm max}$, and for $L(L-D+2) \gg 1 + \eta_{max}^2$ we may use the WKB approximation (\ref{eq:WKB}).  Since the phase $\Upsilon(\eta)$ in (\ref{eq:WKB}) is highly oscillatory at large $L$, we expect $\chi_{\s_1\s_2}(L)$ to decrease rapidly with $L$.  To verify that this is the case, suppose that in fact $L(L-D+2) \gg M^2 \ell^2 (1 + \eta_{max}^2)$ so that we may also expand the phases $\Upsilon_\sigma(\eta)$ (where we have added the label $\sigma$ to indicated the dependence on mass) as
\begin{equation}
\Upsilon_\sigma = \sum_{n \ge 0} \left( L(L-D+2) \right)^{(1-n)/2} \Upsilon_{n,\sigma}.
\end{equation}
Noting that $\Upsilon_{0,\sigma}$ is independent of $\sigma$, we now introduce a new time coordinate $\tilde \eta$ defined by i) $\tilde \eta$ is a smooth strictly increasing function of $\eta$,  ii) $\tilde \eta = \Upsilon_{0,\sigma}$ in the region where $g'(\eta)\neq 0$, and iii) $\tilde \eta = \eta$ at large $|\eta|$.  Then  for large $L$  (\ref{eq:chi}) becomes ($L^{-1}$ times) the Fourier transform of a smooth $L$-independent function of $\tilde \eta$ and, as a result, decays faster than any power law in $L$.   (Here we use the fact that the sub-leading terms in the WKB expansion correct (\ref{eq:WKB}) by multiplying (\ref{eq:WKB}) by functions which become essentially constant at large $L$ for $|\eta| < \eta_{max}$.)   For later use we note that for $L \gg |\eta_{max}| M \ell$ we have shown that $|\chi_{\sigma_1,\sigma_2}(L)| \le 1/L^n$ and, as a result, that for all $L$ and $n$ we have a bound
\begin{equation}
\label{eq:fb}
|\chi_{\s\s}(L)| \le C_n \left( \frac{M \ell}{L}\right)^n
\end{equation}
for appropriate $C_n$ determined by only $g(x)$ (and in particular, for which the $C_n$ do not depend on $M, L$).

It follows that the sum over $L$ in (\ref{eq:T2again}) is absolutely
convergent and yields a finite result, even when $x_1=x_2$.
To see this, we bound the harmonics
$|u^*_{\s L}(x_1) u^*_{\s L}(\eta_2)| \le |u_{\s L}(0)|^2$ as in
(\ref{eq:ubound}), and we bound the Gegenbauer polynomial by it's value
at coincidence:
\eq{
  C^{(D-2)/2}_L(1) = \GGG{L+D-2}{L+1, D-2} ,
}
(see eq. (44) of \cite{Marolf:2008hg}).
For $L \gg 1$ this behaves like $L^{D-3}$.
From these bounds it follows that the summand in (\ref{eq:T2again})
may be bounded at large $L$ by $L^{D-3} \chi_{\s_1\s_2}(L)$. Since
$\chi_{\s_1\s_2}(L)$ decays faster than
any polynomial as $L\to \infty$, the sum is absolutely convergent.

Furthermore, we can show that when $|\eta_1|,\,|\eta_2| \gg |\eta|_{\rm max}^2$
the expression $T_{2\,\s\s}(x_1,x_2)$ decays like $(\eta_1\eta_2)^\s$.
For $\eta_1$, $\eta_2$ in this regime, let us choose some $L_{\rm cut}$
such that
${\rm min}(|\eta_1|, |\eta_2|) \gg 4 L^2_{\rm cut} \gg |\eta|_{\rm max}^2$.
We then split the sum over $L$ in (\ref{eq:T2again}) into two parts: One
part $T_{2,<}(x_1,x_2)$ is the finite series containing the terms below
$L_{\rm cut}$. The other part $T_{2,>}(x_1,x_2)$ is the infinite series
containing the terms with $L \ge L_{\rm cut}$.
For each term in $T_{2,<}(x_1,x_2)$ we may approximate $u^*_{\s L}(\eta_1)$
and $u^*_{\s L}(\eta_2)$ by the asymptotic expansion
(\ref{eq:uLargeEta}).
It follows that each term in the series $T_{2,<}(x_1,x_2)$ decays like
$(\eta_1\eta_2)^\s$, and so $T_{2,<}(x_1,x_2)$ does as well.

On the other hand, we can bound the contribution of the infinite series $T_{2,>}(x_1,x_2)$ by
\eqn{ \label{eq:T2>}
  T_{2>},(x_1,x_2) &\le &
  \ell^{2-D} \frac{\G{\frac{D-2}{2}}}{2\pi^{D/2}}
  \sum_{L=L_{\rm cut}}^{\infty} \bigg\{ (2 L+D-2) |\chi_{\s_1\s_2}(L)|
  |u_{\s L}(0)|^2 C^{(D-2)/2}_L(1)
  \bigg\} .
}
This series converges absolutely and (due to the rapid decay of $\chi_{\s_1\s_2}(L)$) decreases faster than any power law
as $L_{\rm cut}$ is increased. Thus, as $|\eta_{1,2}| \to \infty$
we may increase $L_{\rm cut}$ (taking $L_{\rm cut}$ to be, e.g., any geometric mean of $|\eta|_{max}$ and $\min(|\eta_1|,|\eta_2|)$) and the contribution due to
$T_{2,>}(x_1,x_2)$ becomes negligible. We conclude that the full
expression $T_{2\,\s,\s}(x_1,x_2)$ decays like $(\eta_1\eta_2)^\s$.

For later use, let us also consider $T_{2\,\s,\mu}(x_1,x_2)$ for all $\Re \mu < \Re \sigma$. Of course, the same argument shows that each $T_{2\,\s,\mu}(x_1,x_2)$ decays in the same way, and in particular that it is bounded by $C|(\eta_1\eta_2)^\s|$.
Using (\ref{eq:fb}) we see that one can choose the constant $C$ to grow with $\mu$ at most as some polynomial whose order depends on $D$ through the power $n$ in (\ref{eq:fb}) required to show convergence of the mode sums.

To summarize, the $O(g)$ correction to the time-ordered 2-point function may
be written
\eq{
  \CP{T \phi_\s(x_1) \phi_\s(x_2)}^{(1)}
  = - g_f \d_{M^2} G_{\s}(x_1,x_2) + T_{2\,\s,\s}(x_1,x_2) .
}
The first term is precisely the $O(g)$ correction to the Hartle-Hawking
state \cite{Marolf:2010zp}.
At sufficiently late times $|\eta_{1,2}| \gg |\eta|^2_{\rm max}$
the second term decays like $|\eta_1\eta_2|^\s$; i.e., exponentially in the usual global time coordinate $t$.  Furthermore, the coefficient of the decay term grows at large $\sigma$ no faster than a power law in $\s$, and similarly for all $ T_{2\,\s,\mu}(x_1,x_2)$ with $\Re \mu \le \Re \sigma$.

\subsubsection{The $O(g^2)$ correction}

Let us now compute the $O(g^2)$ correction to the time-ordered 2-point
function. Although it is somewhat tedious, the effort will be worthwhile as we will make use of this result in studying the 1-loop $\phi^3$ correction in the next section.  The $O(g^2)$ correction is depicted in
Fig.~\ref{fig:phi3}~(b) and is given by the expression
\eqn{
  & &\CP{T \phi_{\s_1}(x_1)\phi_{\s_2}(x_2)}^{(2)}
  \nn \\
  &=& i^2 \int_{y} \int_{\yb} g(y) g(\yb) \bigg\{
  G_{\s_1}(y,x_1) G_{\s_2}(\yb,x_2) G_{\s_3}(y,\yb)
  -  W_{\s_1}(y,x_1) G_{\s_2}(\yb,x_2) W_{\s_3}(y,\yb)
  \nn \\ & & \phantom{i^2 \int_{y} \int_{\yb} g(y) g(\yb) \bigg\{ }
  +  W_{\s_1}(y,x_1) W_{\s_2}(\yb,x_2) G^*_{\s_3}(y,\yb)
  -  G_{\s_1}(y,x_1) W_{\s_2}(\yb,x_2) W_{\s_3}(\yb,y)
  \bigg\} . \nn \\
}
This expression may be organized into two terms, each of which contains the $O(g)$ corrections to an appropriate 2-point function:
\eqn{ \label{eq:A0}
  \CP{T \phi_{\s_1}(x_1)\phi_{\s_2}(x_2)}^{(2)}
  &=& i \int_{\yb} g(\yb) G_{\s_2}(\yb,x_2)
  \CP{T \phi_{\s_1}(x_1)\phi_{\s_3}(\yb)}^{(1)}
  \nn \\ & &
  - i \int_{\yb} g(\yb) W_{\s_2}(\yb,x_2)
  \CP{\phi_{\s_3}(\yb)\phi_{\s_1}(x_1)}^{(1)} .
}

The $O(g)$ correction to the time-ordered correlator (first line of (\ref{eq:A0})) was studied in section \ref{sec:Og} above, and the $O(g)$ correction to the Wightman correlator (second line of (\ref{eq:A0}) can be analyzed similarly.  In particular, using manipulations analogous to those that led to (\ref{eq:phi2b}), one obtains
\eqn{ \label{eq:A2}
  & &\CP{\phi_{\s_1}(x_1)\phi_{\s_2}(x_2)}^{(1)}
  = i \int_y g(y) \left[ W_{\s_1}(x_1,y) G_{\s_2}(x_2,y)
    - G_{\s_1}^*(y,x_1) W_{\s_2}(y,x_2) \right]
  \nn \\ & &
  = \frac{1}{M_1^2-M_2^2}\left[ g(x_1) W_{\s_2}(x_1,x_2)
    - g(x_2) W_{\s_1}(x_1,x_2)\right]
  \nn \\ & & \phantom{= }
  + \frac{i}{M_1^2-M_2^2} \int_y(\nabla^\mu g(y))
  \left[
    W_{\s_1}(x_1,y) \KG_\mu G_{\s_2}(x_2,y)
    - G^*_{\s_1}(y,x_1) \KG_\mu W_{\s_2}(y,x_2)
  \right] .
}
After inserting (\ref{eq:phi2c}) and (\ref{eq:A2}) into (\ref{eq:A0})
and rearranging terms one can again recognize the $O(g)$ corrections which
we have already computed:
\eqn{ \label{eq:A3}
  & &\CP{T \phi_{\s_1}(x_1) \phi_{\s_2}(x_2)}^{(2)}
  \nn \\ &=&
  \frac{1}{M_1^2-M_3^2} \left\{
    g(x_1) \CP{T \phi_{\s_3}(x_1)\phi_{\s_2}(x_2)}^{(1)}
    - \CP{T\phi_{\s_1}(x_1)\phi_{\s_2}(x_2)}^{(1')} \right\}
  \nn \\ & &
  + \frac{i}{M_1^2 - M_3^2} \int_y (\nabla^\mu g(y))
  \bigg\{
  - \CP{T\phi_{\s_3}(y)\phi_{\s_2}(x_2)}^{(1)}\KG_\mu G_{\s_1}(y,x_1)
  \nn \\ & & \phantom{+\frac{i}{M_1^2-M_3^2} \int_y (\nabla^\mu g(y))\bigg\{}
    + \CP{\phi_{\s_3}(y)\phi_{\s_2}(x_2)}^{(1)}\KG_\mu W_{\s_1}(y,x_1)
  \bigg\} .
}
Here $\CP{T\phi_{\s_1}(x_1)\phi_{\s_2}(x_2)}^{(1')}$ denotes the same
integral expression (\ref{eq:phi2}) as
$\CP{T\phi_{\s_1}(x_1)\phi_{\s_2}(x_2)}^{(1)}$ but
with a $g(y)$ replaced by $g^2(y)$.
Once again we may use (\ref{eq:phi2c}) and (\ref{eq:A2}) to simplify
this expression. The result may be written
\eqn{ \label{eq:phi2gg}
  \CP{T \phi_{\s_1}(x_1)\phi_{\s_2}(x_2)}^{(2)}
  &=& g^2_f \bigg[
  \frac{G_{\s_1}(x_1,x_2)}{(M_1^2-M_2^2)(M_1^2-M_3^2)}
  + \frac{G_{\s_2}(x_1,x_2)}{(M_2^2-M_1^2)(M_2^2-M_3^2)}
  \nn \\ & & \phantom{g^2 \bigg[}
  + \frac{G_{\s_3}(x_1,x_2)}{(M_3^2-M_1^2)(M_3^2-M_2^2)}
  \bigg] + T_{3\,\s_1\s_2\s_3}(x_1,x_2) ,
}
with $T_{3\,\s_1\s_2\s_3}(x_1,x_2)$ the collection of integration terms
\eq{ \label{eq:T3}
  T_{3\,\s_1\s_2\s_3}(x_1,x_2)
  = (g\nabla g {\rm \; terms}) + (\nabla g \overline{\nabla} g {\rm \; terms}) ,
}
where
\eqn{ \label{eq:gdelg}
  (g \nabla g {\rm \; terms})
  &:=& \frac{1}{(M_1^2-M_3^2)} i \int_y
  \nn \\ & &
    \bigg\{ (\nabla^\mu g(y)) g(x_1)
    \left[\frac{ G_{\s_3}(y,x_1)\KG_\mu G_{\s_2}(y,x_2)
        - W_{\s_3}(y,x_1)\KG_\mu W_{\s_2}(y,x_2)}{(M_3^2-M_2^2)} \right]
    \nn \\ & & \phantom{\bigg\{\;}
    + (\nabla^\mu g^2(y))
    \left[ \frac{ G_{\s_1}(y,x_1)\KG_\mu G_{\s_2}(y,x_2)
        - W_{\s_1}(y,x_1)\KG_\mu W_{\s_2}(y,x_2)}{(M_2^2-M_1^2)} \right]
    \nn \\ & & \phantom{\bigg\{\;}
    + (\nabla^\mu g(y)) g(y)
    \left[\frac{ G_{\s_1}(y,x_1)\KG_\mu G_{\s_2}(y,x_2)
        - W_{\s_1}(y,x_1)\KG_\mu W_{\s_2}(y,x_2)}{(M_3^2-M_2^2)} \right]
    \nn \\ & & \phantom{\bigg\{\;}
    + (\nabla^\mu g(y)) g(x_2)
    \left[ \frac{ G_{\s_1}(y,x_1)\KG_\mu G_{\s_3}(y,x_2)
        - W_{\s_1}(y,x_1)\KG_\mu W_{\s_3}(y,x_2)}{(M_2^2-M_3^2)} \right]
    \bigg\} ,
    \nn \\
}
\eqn{ \label{eq:delgdelg}
  (\nabla g \overline{\nabla}g \;{\rm terms})
  &:=& \frac{1}{(M_1^2-M_3^2)(M_2^2-M_3^2)}
  \int_y \int_\yb (\nabla^\mu g(y))(\nabla^{\bnu}g(\yb))
  \nn \\ & &
  \bigg\{ W_{\s_1}(x_1,y) \KG_\mu G_{\s_3}(\yb,y) \KG_{\bnu} W_{\s_2}(x_2,\yb)
 \nn \\ & & \phantom{\bigg\{ \;}
 - W_{\s_1}(x_1,y) \KG_\mu W_{\s_3}(\yb,y) \KG_{\bnu} W_{\s_2}(\yb,x_2)
  \nn \\ & & \phantom{\bigg\{ \;}
  - W_{\s_1}(y,x_1) \KG_\mu W_{\s_3}(y,\yb) \KG_{\bnu} W_{\s_2}(x_2,\yb)
 \nn \\ & & \phantom{\bigg\{ \;}
  + W_{\s_1}(y,x_1) \KG_\mu G^*_{\s_3}(y,\yb) \KG_{\bnu} W_{\s_2}(\yb,x_2)
  \bigg\}  .
}

Let us simplify the lengthy expressions (\ref{eq:gdelg}) and
(\ref{eq:delgdelg}). The terms in (\ref{eq:gdelg}) are of the same form as
$T_{2\,\s_1\s_2}(x_1,x_2)$ above; indeed, after a few simple manipulations we
may write (\ref{eq:gdelg}) as
\eqn{
  (g \nabla g {\rm \; terms})
  &=& \frac{g_f}{(M_1^2 - M_3^2)} T_{2\,\s_3,\s_2}(x_1,x_2)
  + \frac{g_f}{(M_2^2 - M_3^2)} T_{2\,\s_1,\s_3}(x_1,x_2)
  \nn \\ & &
  + \frac{(M_1^2+M_2^2-2M_3^2)}{2(M_3^2-M_2^2)(M_1^2-M_3^3)}
  T_{2'\,\s_1\s_2}(x_1,x_2)  .
}
Here $T_{2'\,\s_1\s_2}(x_1,x_2)$ denotes the same integral expression
(\ref{eq:T2}) as $T_{2\,\s_1\s_2}(x_1,x_2)$ but with $g^2(y)$ in place of $g(y)$.
The $(\nabla g \overline{\nabla} g {\rm \; terms})$ may be decomposed
into the sum of the two terms
\eqn{ \label{eq:TW}
  & & T_{W\,\s_1\s_2,\s_3}(x_1,x_2)
  := \frac{-2}{(M_1^2-M_3^2)(M_2^2-M_3^2)}
  \nn \\ & & \;\;\;
  \Re \left\{ \int_y \int_\yb (\nabla^\mu g(y))(\nabla^{\bnu}g(\yb))
  W_{\s_1}(x_1,y) \KG_\mu W_{\s_3}(\yb,y) \KG_{\bnu} W_{\s_2}(\yb,x_2) \right\} ,
  \\
\label{eq:TG}
  & & T_{G\,\s_1\s_2,\s_3}(x_1,x_2)
  := \frac{2}{(M_1^2-M_3^2)(M_2^2-M_3^2)}
  \nn \\ & &  \;\;\;
  \Re \left\{ \int_y \int_\yb (\nabla^\mu g(y))(\nabla^{\bnu}g(\yb))
  W_{\s_1}(x_1,y) \KG_\mu G_{\s_3}(\yb,y) \KG_{\bnu} W_{\s_2}(x_2,\yb) \right\}.
}
Expanding the Green's functions in modes, one easily obtains
\eqn{ \label{eq:TW2}
  & & T_{W\,\s_1\s_2,\s_3}(x_1,x_2)
  \nn \\
  &=&
  \ell^{2-D} \frac{\G{\frac{D-2}{2}}}{2\pi^{D/2}}
  \Re \left\{
  \sum_{L=0}^\infty (2L+D-2) \chi_{\s_1 \s_3}^*(L) \chi_{\s_3 \s_2}(L)
  u_{\s_1 L}(\eta_1) u^*_{\s_2 L}(\eta_2) C_L^{(D-2)/2}(\vx_1\cdot\vx_2)
\right\}. \nn \\
}

This leaves only $T_{G\,\s_1\s_2\s_3}(x_1,x_2)$, which one may treat similarly
using
$G_\sigma(\bar y,y) = \theta(\bar \eta - \eta) W_\sigma(\bar y, y) +\theta(\bar \eta - \eta) W^*_\sigma(\bar y, y) $
and the mode sum (\ref{eq:Wmodes}). One finds
\eqn{ \label{eq:TG2}
  & & T_{G\,\s_1\s_2,\s_3}(x_1,x_2)
  \nn \\
  &=&
  \ell^{2-D} \frac{\G{\frac{D-2}{2}}}{2\pi^{D/2}(M_2^2-M_3^2)}
  \Re \Bigg\{
  \sum_{L=0}^\infty   C_L^{(D-2)/2}(\vx_1\cdot\vx_2) u_{\s_1 L}(\eta_1) u_{\s_2 L}(\eta_2) \cr &\times& \frac{1}{L^2}\left( \int d \bar \eta (1+\bar \eta^2)^{(D-1)/2} g'(\bar \eta) [u^*_{\sigma_3 L} (\bar \eta) \overleftrightarrow{\d_n} u^*_{\sigma_2 L} (\bar \eta)] \zeta_{L \sigma_1 \sigma_3} (\bar \eta) + (\sigma_1 \leftrightarrow \s_2) \right)
\Bigg\}, \nn \\
}
where
\eq{ \label{eq:zeta}
  \zeta_{L \s_1\s_3}(\bar \eta)
  = \frac{L^2}{(M_1^2-M_3^2)} \int_{\bar \eta}^\infty d\eta\, (1+\eta^2)^{(D-1)/2}
  g'(\eta) \left[
    u^*_{\s_1 L}(\eta) \overleftrightarrow{\d_n}
    u_{\s_3 L}(\eta) \right] .
}
Note that the integral in (\ref{eq:zeta}) converges since the integrand is smooth and $g'(\eta)$ has compact support.  Using (\ref{eq:WKB}), we see that the large $L$ limit of (\ref{eq:zeta}) is a smooth function of $\bar \eta$ which is in fact independent of $L,M_1,M_3.$ As a result, the $\bar \eta$ integral in (\ref{eq:TG2}) once again gives a function of $L$ which decreases faster than any power of $L$.  In particular, as with $\chi_{\s_1 \s_2}(L)$, for any $p > 0$ it may be bounded by $C_p(\s_1,\s_3) L^{-p}$ where $C_p(\s_1,\s_3)$ is a polynomial in $\s_1,\s_3$ whose order and coefficients are determined by $p$.

It is clear that (\ref{eq:TW2}) and (\ref{eq:TG2}) are similar in form to (\ref{eq:T2}).  As a result,  $T_{W\,\s_1\s_2\s_3}(x_1,x_2)$ and $T_{G\,\s_1\s_2\s_3}(x_1,x_2)$ can be shown to decay like $\eta_1^{\sigma_1}\eta_1^{\sigma_2}$ via arguments analogous to those used for $T_{2\,\s_1\s_2}(x_1,x_2)$.  
Furthermore, just as with $T_{2\,\s_1\s_2}(x_1,x_2)$, we find that these 
functions are in fact bounded by $C |(\eta_1)^\sigma_1(\eta_1)^\sigma_2|$ for some polynomial function $C(\s_1,\s_2)$.

Collecting our results, we conclude that the $O(g^2)$ correction to
the time-ordered 2-point function is
\eqn{ \label{eq:result}
  & & \CP{T \phi_{\s_1}(x_1)\phi_{\s_2}(x_2)}^{(2)}
  \nn \\
  &=& g^2_f \bigg[
  \frac{G_{\s_1}(x_1,x_2)}{(M_1^2-M_2^2)(M_1^2-M_3^2)}
  + \frac{G_{\s_2}(x_1,x_2)}{(M_2^2-M_1^2)(M_2^2-M_3^2)}
  + \frac{G_{\s_3}(x_1,x_2)}{(M_3^2-M_1^2)(M_3^2-M_2^2)}
  \bigg]
  \nn \\ & &
  + \frac{g_f}{(M_1^2 - M_3^2)} T_{2\,\s_3,\s_2}(x_1,x_2)
  + \frac{g_f}{(M_2^2 - M_3^2)} T_{2\,\s_1,\s_3}(x_1,x_2)
  \nn \\ & &
  + \frac{(M_1^2+M_2^2-2M_3^2)}{2(M_3^2-M_2^2)(M_1^2-M_3^3)}
  T_{2'\,\s_1\s_2}(x_1,x_2)
  + T_{W\,\s_1\s_2,\s_3}(x_1,x_2)
  + T_{G\,\s_1\s_2,\s_3}(x_1,x_2) . \nn \\
}
We remind the reader that $T_{2\,\s_1\s_2}(x_1,x_2)$ is defined in
(\ref{eq:T2}), $T_{2'\,\s_1\s_2}(x_1,x_2)$ is (\ref{eq:T2}) with
$g^2(y)$ in place of $g(y)$,
$T_{W\,\s_1\s_2,\s_3}(x_1,x_2)$ is defined in (\ref{eq:TW}), and
$T_{G\,\s_1\s_2,\s_3}(x_1,x_2)$ is defined in (\ref{eq:TG}).
In (\ref{eq:result}), the term in square brackets is the $O(g^2)$ correction
to the Hartle-Hawking state. The remaining terms are bounded
for all $x_1,x_2 \in J^+(\cR)/\cR$ and in particular are finite at coincidence ($x_1=x_2$).  In addition, these terms all decay like
$\eta_1^{\s_1} \eta_2^{\s_2}$ when $|\eta_{1,2}| > |\eta|_{\rm max}$.
One may readily verify that the full expression (\ref{eq:result})
is regular in the limit of coincident masses. In this limit, and at
late times, we obtain
\eq{
  \CP{T \phi_{\s}(x_1)\phi_{\s}(x_2)}^{(2)}
  = \half g_f^2 \d^2_{M^2} G_{\s}(x_1,x_2) + O\left((\eta_1\eta_2)^\s\right) .
}

\subsection{Example: $\phi^3(x)$ interaction}
\label{sec:phi3}

\begin{figure}
  \centering
  \includegraphics{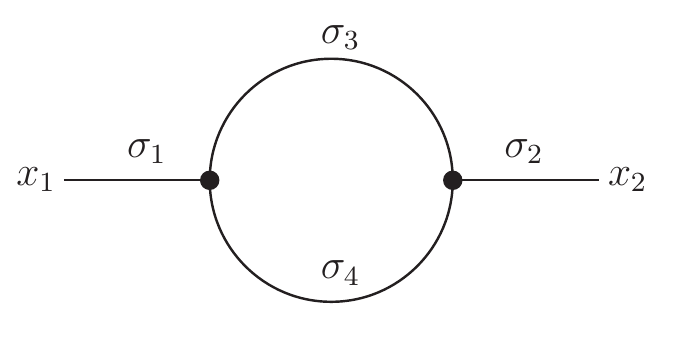}
  \caption{The $O(g^2(x))$ correction to the time-ordered 2-point
    function $\CP{T\phi(x_1)\phi(x_2)}$ in a theory with an
    $g(x)\phi^3(x)$ interaction. Once again we label each
    leg of the diagram by a distinct mass parameter $\s_i$.}
  \label{fig:phi3}
\end{figure}
\renewcommand{\thefigure}{\arabic{figure}}

In this section we consider a cubic self-interaction
\eq{
  \cL_{\rm int}[\phi] = - \frac{g(x)}{3!}\phi^3(x) ,
}
and compute the $O(g^2)$ correction to the time-ordered 2-point
function. The relevant Feynman diagram is shown in Fig.~\ref{fig:phi3}.
In spacetime dimension $D\ge 4$ this correction contains ultraviolet
divergences, so we will need to regulate our computation and renormalize
the theory. We restrict attention to $D \le 6$ for which
$\phi^3$-theory is power-counting normalizable; for these dimensions,
and to $O(g^2)$ the counterterms needed for renormalization are
\eq{ \label{eq:Lct}
  \cL_{\rm ct}[\phi]
  = -\frac{(Z_\phi(x)-1)}{2} \nabla_\mu\phi(x)\nabla^\mu\phi(x)
  - \frac{(Z_M(x)-1)M^2}{2}\phi^2(x) ,
}
with $Z_\phi(x)$ and $Z_M(x)$ given by $Z_i = 1 + O(g^2)$. The
renormalization coefficients are position-dependant as a result of
our position-dependant coupling $g(x)$. However, no renormalization is
required for $D=2,3$.

To regulate
our computation of the diagram (Fig.~\ref{fig:phi3}) we replace the
internal Green's functions with Pauli-Villars regulated Green's functions.
Thus the full $O(g^2)$ correction is given schematically by
\eq{ \label{eq:phi32pt}
  \CP{T \phi_{\s_1}(x_1)\phi_{\s_2}(x_2)}^{(2)} = ({\rm diag}) + ({\rm c.t.}) ,
}
where $({\rm diag})$ is the regulated Feynman diagram
given by the expression
\eqn{ \label{eq:diag}
  ({\rm diag}) &=&
  i^2 \int_{y_1} \int_{y_2} g(y_1) g(y_2)
  \bigg\{
  G_{\s_1}(y_1,x_1) G_{\s_2}(y_2,x_2)
  G^{\rm reg}_{\s_3}(y_1,y_2) G^{\rm reg}_{\s_4}(y_1,y_2)
  \nn \\ & & \phantom{i^2 \int_{y_1} \int_{y_2} g(y_1) g(y_2)\bigg\{}
  -  W_{\s_1}(y_1,x_1) G_{\s_2}(y_2,x_2)
  W^{\rm reg}_{\s_3}(y_1,y_2) W^{\rm reg}_{\s_4}(y_1,y_2)
  \nn \\ & & \phantom{i^2 \int_{y_1} \int_{y_2} g(y_1) g(y_2)\bigg\{}
  +  W_{\s_1}(y_1,x_1) W_{\s_2}(y_2,x_2)
  G^{{\rm reg}\,*}_{\s_3}(y_1,y_2) G^{{\rm reg}\,*}_{\s_4}(y_1,y_2)
  \nn \\ & & \phantom{i^2 \int_{y_1} \int_{y_2} g(y_1) g(y_2)\bigg\{}
  -  G_{\s_1}(y_1,x_1) W_{\s_2}(y_2,x_2)
  W^{\rm reg}_{\s_3}(y_2,y_1) W^{\rm reg}_{\s_4}(y_2,y_1)
  \bigg\} ,
}
and $(\rm c.t.)$ are the counterterms generated by (\ref{eq:Lct}):
\eqn{ \label{eq:cts}
  ({\rm c.t.}) &=&  - (Z_\phi(x_2)-1) G_{\s_1}(x_1,x_2)
  \nn \\ & &
  + i M_2^2
  \int_y \left(Z_\phi(y) + Z_M(y) - 2 \right)
  \left[ G_{\s_1}(y,x_1) G_{\s_2}(y,x_2)
   - W_{\s_1}(y,x_1) W_{\s_2}(y,x_2) \right] .\quad
}
At the end of our computation we will set the masses to be equal.

A useful way to proceed is to make use of the linearization formulae
for Bunch-Davies Green's functions:
\eq{ \label{eq:lin}
  H^{\rm reg}_{\s_1}(x,y)H^{\rm reg}_{\s_2}(x,y)
  = \int_\mu f(\mu) H_\mu(x,y) .
}
Here $H_\s(x,y)$ and $H^{\rm reg}_\s(x,y)$ may be taken to be any
Bunch-Davies Green's function (e.g., time-ordered, Wightman, etc.). In addition to $\mu$, the function
$f(\mu)$ implicitly depends upon $\s_1$, $\s_2$, and the spacetime
dimension, as well as the collection of Pauli-Villars masses for
the two Green's functions. To derive an expression for the function
$f(\mu)$ it is sufficient to construct the linearization formula
for the Euclidean Green's function $\D_\s(x,y)$. Since the Bunch-Davies
Green's functions are given by the Euclidean Green's function with
an appropriately chosen prescription for avoiding the cut in the
complex $Z$ plane, the extension of the linearization formula to
these Green's functions follows immediately. We derive the linearization
formulae in Appendix~\ref{app:linearization} following \cite{Marolf:2010nz}; here we will simply state the
results. For the time-ordered Green's function we have
\eqn{ \label{eq:Glin}
  G^{\rm reg}_{\s_1}(x,y)G^{\rm reg}_{\s_2}(x,y)
  &=& \ell^{2-D} \int_\mu
  \left(f_{\s_1\s_2}(\mu)+f^{\rm van}(\mu)\right) G_\mu(x,y) ,
  \quad\quad\quad\quad\quad\quad\quad\quad\quad
  (D=2,3) \nn \\
  G^{\rm reg}_{\s_1}(x,y)G^{\rm reg}_{\s_2}(x,y)
  &=& \ell^{2-D}
  \int_\mu \left(f_{\s_1\s_2}(\mu)+f^{\rm van}(\mu)\right) G_\mu(x,y)
  - i \ell^{4-D} c_0 \delta(x,y) ,
  \quad\quad
  (D=4,5) \nn \\
  G^{\rm reg}_{\s_1}(x,y)G^{\rm reg}_{\s_2}(x,y)
  &=& \ell^{2-D}
  \int_\mu \left(f_{\s_1\s_2}(\mu)+f^{\rm van}(\mu)\right) G_\mu(x,y)
  \nn \\ & &
  -i \ell^{4-D} c_0 \delta(x,y)
  -i \ell^{6-D} c_1 \Box \delta(x,y) , \quad\quad (D=6,7) .
}
where $f_{\s_1\s_2}(\mu)$ depends only on $\sigma_1,\sigma_2$ and is independent of the regulator masses.
In the complex $\mu$ plane the function
$(f_{\s_1\s_2}(\mu)+ f^{\rm van}(\mu))$ decays exponentially
away from the real axis and is analytic in the strip
$\Re(\s_1+\s_2) < \Re \mu$. The contours of integration
(\ref{eq:Glin}) and (\ref{eq:Wlin}) lie within the strip
$\Re(\s_1+\s_2) < \Re \mu < 0$. 

The coefficients $c_0$ and $c_1$ in (\ref{eq:Glin}) are real functions
of the Pauli-Villars masses but do not depend on $x,y$ or the integration
variable $\mu$.
These expressions have been organized to make it easy to take the limit of large
Pauli-Villars regulator masses.
As discussed in the Appendix, in this limit the function
$f_{\s_1\s_2}(\mu)$ remains, $f^{\rm van}(\mu)$ vanishes, and
$c_0$ and $c_1$ diverge. The explicit expressions
for $f_{\s_1\s_2}(\mu)$, $f^{\rm van}(\mu)$, $c_0$ and $c_1$ can
be found in the Appendix, but we will not need them here.
For the Wightman Green's function we have
\eq{ \label{eq:Wlin}
  W^{\rm reg}_{\s_1}(x,y)W^{\rm reg}_{\s_2}(x,y)
  = \ell^{2-D} \int_\mu (f_{\s_1\s_2}(\mu)+ f^{\rm van}(\mu)) W_\mu(x,y) ,
  \quad ({\rm all}\; D) .
}
This expression is finite when the Pauli-Villars masses are taken
to infinity. This reflects that fact that the product of such
Wightman functions
$W_{\s_1}(x,y)\cdots W_{\s_n}(x,y)$ is positive/negative frequency
with respect to $x$/$y$.

Let us start with the simple $D=2,3$ where there are no
ultraviolet divergences at this order.  In this case we may set
$Z_M = Z_\phi = 1+O(g^4)$ and immediately take the limit of large regulator masses\footnote{We could also have done the full computation without ever involving regulators.  The unregulated Green's functions for $D=2,3$ satisfy the analogues of (\ref{eq:Glin}), (\ref{eq:Wlin}) with $f^{van} =0$.} (which sends $f^{\rm van}(\mu)$ to zero)
in the linearization formulae (\ref{eq:Glin}) and (\ref{eq:Wlin}).
The full correction to the 2-point function
(\ref{eq:phi32pt}) then becomes
\eqn{  \label{eq:phi3tree}
  & & \CP{T \phi_{\s_1}(x_1)\phi_{\s_2}(x_2)}^{(2)}
  = \ell^{2-D} \int_\mu f_{\s_3\s_4}(\mu)
  \bigg\{
  \nn \\ & &
  i^2 \int_{y_1} \int_{y_2} g(y_1) g(y_2)
  \bigg[
  G_{\s_1}(y_1,x_1) G_{\s_2}(y_2,x_2)
  G_{\mu}(y_1,y_2)
  -  W_{\s_1}(y_1,x_1) G_{\s_2}(y_2,x_2)
  W_{\mu}(y_1,y_2)
  \nn \\ & & \phantom{i^2 \int_{y_1} \int_{y_2} g(y_1) g(y_2)
  \bigg\{ }
  +  W_{\s_1}(y_1,x_1) W_{\s_2}(y_2,x_2)
  G^*_{\mu}(y_1,y_2)
  -  G_{\s_1}(y_1,x_1) W_{\s_2}(y_2,x_2)
  W_{\mu}(y_2,y_1)
  \bigg]\bigg\} . \nn \\
}
The astute reader will recognize the term in braces as the expression
for the $O(g^2)$ correction to the 2-point function in the $\phi^2$-theory
discussed in \S\ref{sec:phi2}. Inserting our final expression (\ref{eq:result})
for that correction we obtain
\eqn{
  \label{eq:int}
  & & \CP{T \phi_{\s_1}(x_1)\phi_{\s_2}(x_2)}^{(2)} =
  \nn \\ & &
  \ell^{2-D} \int_\mu f_{\s_3\s_4}(\mu) \bigg\{
  g^2_f \bigg[
  \frac{G_{\s_1}(x_1,x_2)}{(M_1^2-M_2^2)(M_1^2-M_\mu^2)}
  + \frac{G_{\s_2}(x_1,x_2)}{(M_2^2-M_1^2)(M_2^2-M_\mu^2)}
  + \frac{G_{\mu}(x_1,x_2)}{(M_\mu^2-M_1^2)(M_\mu^2-M_2^2)}
  \bigg]
  \nn \\ & & \phantom{  \ell^{-1} \int_\mu f_{\s_1\s_2}(\mu) \bigg\{}
  + T_{3\,\s_1\s_2\mu}(x_1,x_2) \bigg\}
  \\ &=&
  \ell^{2-D} \int_\mu f_{\s_3\s_4}(\mu) \bigg\{
  g^2_f \frac{G_{\mu}(x_1,x_2)}{(M_\mu^2-M_1^2)(M_\mu^2-M_2^2)}
  + g_f \frac{T_{2\,\mu \s_2}(x_1,x_2)}{(M_1^2 - M_\mu^2)}
  + g_f \frac{T_{2\,\s_1 \mu}(x_1,x_2)}{(M_2^2 - M_\mu^2)}
  \nn \\ & & \phantom{\ell^{-1} \int_\mu f_{\s_3\s_4}(\mu) \bigg\{\;\;}
  + T_{W\,\s_1\s_2\mu}(x_1,x_2)
  + T_{G\,\s_1\s_2\mu}(x_1,x_2)
  \bigg\},
}
for $x,y \in J^+(\cR)/\cR$. To obtain the second line we have inserted the definition of $T_{3\,\s_1\s_2\mu}(x_1,x_2)$ and noted that terms in the integrand of
(\ref{eq:int}) whose only dependence upon $\mu$ is a factor
of $1/(M_i^2 - M_\mu^2)$ make no contribution to the integral over $\mu$.
This is because $1/(M_i^2 - M_\mu^2)$ contributes only poles to the left
of the integration contour. For these terms the integration contour may
be closed to the right. But there are no poles contained in the right
half-plane, so these integrals vanish.

In higher dimensions the only additional complication is that we
must take care to cancel the ultraviolet divergences contained in
(\ref{eq:diag}) with our available counterterms. Our use of
Pauli-Villars regularization as well as our linearization formulae
make this procedure quite transparent. Consider first the case
of $D=4,5$. After utilizing our linearization
formulae (\ref{eq:Glin}) and (\ref{eq:Wlin}) the divergent terms in
$\CP{T \phi_{\s_1}(x_1)\phi_{\s_2}(x_2)}^{(2)}$ are
\eq{
   + c_0 \ell^{4-D} i \int_y g^2(y) \left[
     G_{\s_1}(y,x_1) G_{\s_2}(y,x_2)
     - W_{\s_1}(y,x_1) W_{\s_2}(y,x_2) \right] .
}
Comparing this expression to our counterterms (\ref{eq:cts}) we see that
these terms are cancelled by setting
\eq{
 Z_M(x) = 1 - \frac{c_0 \ell^{4-D}}{M_2^2} g^2(x) + O(g^4), \quad
 Z_\phi(x) = 1 + O(g^4),
 \quad (D = 4,5) .
}
In $D=6$ dimensions the divergent terms are
\eq{
  - c_1 g_f^2 G_{\s_1}(x_1,x_2)
  + \left(c_0 \ell^{-2}+ c_1 M_2^2 \right) i \int_y g^2(y) \left[
    G_{\s_1}(y,x_1) G_{\s_2}(y,x_2)
    - W_{\s_1}(y,x_1) W_{\s_2}(y,x_2) \right] ,
}
which may be canceled by setting
\eq{
 Z_M(x) = 1 - \left(\frac{c_0}{M_2^2\ell^2} + c_1 \right) g^2(x) + O(g^4),
 \quad
 Z_\phi(x) = 1 - c_1 g_f^2 + O(g^4) ,
 \quad (D = 6) .
}
With all divergent terms cancelled by the counterterms
we may take the Pauli-Villars masses to infinity, so that
$f^{\rm van}(\mu) \to 0$, and then proceed as for $D=2,3$ above.  For all $2 \le D \le 6$ we find
\eqn{ \label{eq:phi3final}
  \CP{T \phi_{\s}(x_1)\phi_{\s}(x_2)}^{(2)} &=&
  \ell^{2-D} \int_\mu f_{\s\s}(\mu) \bigg\{
  g^2_f \frac{G_{\mu}(x_1,x_2)}{(M_\mu^2-M_\s^2)^2}
  + g_f \frac{T_{2\,\mu \s}(x_1,x_2)}{(M_\s^2 - M_\mu^2)}
  + g_f \frac{T_{2\,\s \mu}(x_1,x_2)}{(M_\s^2 - M_\mu^2)}
  \nn \\ & & \phantom{\ell^{-1} \int_\mu f_{\s\s}(\mu) \bigg\{\;\;}
  + T_{W\,\s\s,\mu}(x_1,x_2) + T_{G\,\s\s,\mu}(x_1,x_2)
  \bigg\},
}
where we have now taken the limit in which
all masses become equal.
Note that the units are correct: mass${}^2$ has units $\ell^{-2}$,
$g^2(x)$ has units $\ell^{D-6}$, and $G_\s(x_1,x_2)$, $T_{2\,\s_1\s_2}(x_1,x_2)$
$T_{W\,\s_1\s_2\mu}(x_1,x_2)$, $T_{G\,\s_1\s_2\mu}(x_1,x_2)$ each has
units $\ell^{2-D}$.

The first term in (\ref{eq:phi3final}) gives precisely
$\C{T \phi_{\s}(x_1)\phi_{\s}(x_2)}^{(2)}_{HH,g_f}$, the associated correction to the correlator in the Hartle-Hawking vacuum of the theory with $g=g_f={\rm const.}$ for all time.  Each of the remaining terms inside the braces was analyzed in detail in section \ref{sec:phi2}.  Choosing the $\mu$ contour to satisfy $\Re \mu \le \Re \sigma$, we see that each such term is bounded by $C(\mu)|\eta_1\eta_2|^\sigma$. Recalling that $C$ depends at most polynomially on $\mu$ while $f_{\s\s}$ decays exponentially at large imaginary $\mu$ we see that the integral of these terms over $\mu$ is bounded by $\tilde C |\eta_1\eta_2|^\sigma$ for some constant $\tilde C$; i.e.,
\eq{ \label{eq:phi3finallate}
  \CP{T \phi_{\s}(x_1)\phi_{\s}(x_2)}^{(2)}
  = g_f^2 \ell^{2-D} \int_\mu f_{\s\s}(\mu)
  \frac{G_{\mu}(x_1,x_2)}{(M_\mu^2-M_\s^2)^2}
  + O\left( (\eta_1\eta_2)^\s \right) ,
}
where the first term is precisely the one-loop correction \cite{Marolf:2010nz}
$\C{T \phi_{\s}(x_1)\phi_{\s}(x_2)}^{(2)}_{HH,g_f}$ to the Hartle-Hawking vacuum of the theory with constant $g = g_f$.

We emphasize that the $O\left( (\eta_1\eta_2)^\s \right)$ term is finite at coincidence ($x_1=x_2$).  It follows immediately that the one point function of the composite operator $\phi^2(x_1)$ in our state can again be written as the sum of two terms, the first being its (finite) value in the Hartle-Hawking state for constant $g$ and the other decaying like $\eta_1^{2\s}$.  In particular, defining $\phi^2(x_1)$ via any de Sitter-invariant regulator gives a result of the form

\begin{equation}
\label{eq:phi2res}
\CP{\phi_{\s}^2(x_1)}^{(2)} = {\rm const.} + O((\eta_1)^{2\s}).
\end{equation}

\section{Discussion}
\label{sec:disc}

We have shown that, in the time-dependent model of section \ref{sec:time}, the two-point function at one-loop level approaches that of the constant coupling Hartle-Hawking state in the limit where its arguments are evaluated at late times $\eta_1,\eta_2$.   A key feature of this model is a time-dependent cubic coupling $g(x) \phi^3$ which vanishes before some fixed initial (global) time which we may call $\eta_0$.  We also chose $g(x)$ to be a constant ($g_f$) to the future some Cauchy surface.  The state was taken to coincide with the (free) Bunch-Davies vacuum in the region where $g(x)=0$, so that the state in the region with $g=g_f$ is determined by the particular way in which the coupling turns on.  Since we required only that $g(x)$ be smooth in this transition region, this scenario describes a wide range of possible states.  Our results were established by working entirely in Lorentz signature and in global coordinates.

Our main result is that the difference between our two-point function and that of the $g=g_f$ Hartle-Hawking state is negligible for $\eta_1,\eta_2 \gg |\eta_0|$.  In other words, while the discrepancy may be significant (and perhaps even large!) for some period of time, it decreases rapidly once the universe enters its expanding phase and the size of the spheres becomes significantly larger than they were when the coupling turned on.  This is precisely what one would expect based on the free theory in which perturbations rapidly disperse as, after this time, their effect on local correlators decays rapidly. This behavior was shown in  \cite{Hollands:2010pr,Marolf:2010nz} to hold to all orders in perturbation theory for a dense set of states; our  results here indicate that this dense domain allows for physically interesting initial conditions.

The above model was recently considered by Krotov and Polyakov \cite{Krotov:2010ma}. While they characterized the model as ``unstable,'' we remind the reader that their technical results are completely consistent with ours.  As stated in their paper (below their equation (17)), their analysis applies in the regime $\eta_0 \ll - |\eta_1|, -|\eta_2|$ (our notation); i.e., in precisely the complimentary regime to that studied here.  As noted in the introduction, the divergence they find as $\eta_0 \rightarrow -\infty$ is not only consistent with, but in fact is naturally expected from, the results found here and in \cite{Hollands:2010pr,Marolf:2010nz}.

Despite various technical features of our analysis, we see that the approach to the Hartle-Hawking state at late times followed from a few simple ingredients.  First, for quadratic perturbations $g(x)\phi^2$, one can write the full two-point function at each order as a sum of the Hartle-Hawking two-point function and a `boundary term' associated with the transition region (see (\ref{eq:T1b}), (\ref{eq:A3})).  These manipulations involve only integrations by parts and will clearly hold in a general spacetime which is asymptotically de Sitter to the future.  The rapid expansion of the universe at late times then i) causes any given mode to decay as a power law in $\eta$ and ii) implies that the modes which have not yet decayed at some late time $\eta$  correspond to very high $L$. As a result, at the early time $\eta_0$ when $g(x)$ was time dependent, these high $L$ modes were very high frequency.  Since quadratic perturbations do not lead to loops, the Green's functions that appear in this boundary term are all positive frequency, at least at the key step (see the discussion of $T_{G\,\s_1\s_2,\s_3}(x_1,x_2)$ surrounding (\ref{eq:TG2})). As a result, in the boundary term these modes appear with coefficients involving what is effectively the Fourier transform of $g(x)$ at large momentum, which vanishes rapidly since $g(x)$ is smooth.  Thus the effect of the boundary term decays with time, leaving only the Hartle-Hawking term in the two-point function.

For the cubic perturbation $g(x) \phi^3$, we used the linearization formulae (\ref{eq:lin}) to make renormalization straightforward and to reduce the one-loop calculation to the quadratic-perturbation calculation described above.  It is clear that analogous results follow immediately in any context where similar linearization formulae hold for the associated free Green's functions. While such formulae are not obvious for general fields in general spacetimes, they must hold for conformally coupled free fields in general spherically symmetric spacetimes (which are necessarily conformal to dS), at least after inserting  powers of the appropriate conformal factor $\Omega(x)$.  Note that this argument requires conformal invariance only for the free theory about which we perturb\footnote{Recall that conformally coupled free fields correspond to $\sigma = -(D-2)/2$ in de Sitter.} and that there is no restriction on the interaction.  The only impact of these extra factors of $\Omega(x)$ is to provide what is in effect an extra time-dependence in the coupling. Thus, to the extent that one can study a theory of general spacetime-dependent mass $m^2(x)$ by perturbing the free conformally coupled theory by a quadratic perturbation $g(x) \phi(x)^2$, our cubic results extend for all $M^2 > 0$ to any spherically symmetric spacetime which is asymptotically de Sitter in the far future.

It is useful to comment on the special case where the spacetime is taken to be the Einstein Static Universe (ESU) $S^3 \times {\mathbb R}$ to the past of some $S^3$.  Since the ESU is static and spatially compact, we can take a limit where the coupling $g(x)$ is turned on adiabatically slowly (after which it is $g=g_f = {\rm const.}$) and in the distant past.  We are then guaranteed that, in the ESU region, the state is given by the interacting ESU vacuum.  As a result, subject to the same qualifiers as above we may consider the theory with $g=g_f= {\rm const.}$ for all time.  Taking the state to be the (interacting) ESU vacuum at early times, we see that at late times correlators again approach those of the de Sitter Hartle-Hawking vacuum.  This result can then be further generalized to either $n-$particle or thermal states in the ESU, all of which approach the same de Sitter Hartle-Hawking vacuum at late times.

In summary, we have shown at the one-loop level that a wide class of physically-motivated initial conditions lead to two-point functions which approach that of the Hartle-Hawking state at late times.  This suggests that states defined by general physical initial conditions lie in the dense set of states where the cosmic quantum no hair theorems of \cite{Hollands:2010pr,Marolf:2010nz} apply.
We expect that this can be explicitly checked by extending the calculations reported here to all orders in perturbation theory. After all,  the techniques used above were essentially Lorentz-signature versions of the Euclidean methods applied in \cite{Marolf:2010zp,Marolf:2010nz}. So by adapting further such techniques to Lorentz signature we expect to obtain all-orders results analogous to those found in \cite{Hollands:2010pr,Marolf:2010nz}.

Some readers may feel a lingering uneasiness with these results due to the well-established infrared divergences (see e.g. \cite{Sasaki:1992ux,Polyakov:2007mm,Polyakov:2009nq,Higuchi:2009ew}) associated with in-out perturbation theory in global de Sitter.  In particular, as noted in e.g. \cite{Marolf:2010zp}, at sufficient loop orders such divergences occur in the future expanding region even for very large masses.  This certainly indicates that {\it some} quantity is becoming large in the infrared.  However, the key point to realize is that the quantity need not be local.  In particular, we suggest that it is merely the operator relating the (free) Bunch-Davies vacuum $|0\rangle$ to the interacting Hartle-Hawking state $|HH\rangle$ which becomes large at late times. This operator involves integrals over an entire $S^3$ at each time and can become large as the $S^3$ grows in size.  In Minkowski space, a corresponding IR divergence is forbidden due to the exponential decay of massive propagators at large spacelike separations.  But since the volume element also grows exponentially in dS, such IR divergences can occur.  Indeed, the operator relating $|0\rangle$ and $|HH \rangle$ is closely related to the vacuum to $n$-particle amplitudes of in-out perturbation theory noted to diverge above. This stands in sharp contrast to the good IR behavior of (unintegrated) $n$-point functions as established here and in \cite{Marolf:2010zp,Hollands:2010pr,Marolf:2010nz}.

\vspace{2cm}

\noindent{\bf Acknowledgements:} It is a pleasure to thank Atsushi Higuchi,
Viatcheslav Mukhanov, Alexander Polyakov, Albert Roura, and Richard Woodard
for useful
discussions. DM and IM are supported in part by the US National Science
Foundation under NSF grant PHY08-55415 and by funds from the University
of California.

\appendix

\section{Asymptotic expansion of Klein-Gordon modes}
\label{app:asymptotic}

One way to derive (\ref{eq:uLargeEta}) is to use a standard Laurent
expansion for the Gauss hypergeometric function. Recall that
$F_{\s L}(\eta)$ is a Gauss hypergeometric function
\eq{
  F_{\s L}(\eta) =
  \2F1{\s+\frac{D}{2}}{1 - \s -\frac{D}{2}}{L+\frac{D}{2}}{\frac{1-i\eta}{2}} .
}
For $|(1-i\eta)/2| < 1$ this hypergeometric function has the power series
expansion:
\eqn{ \label{eq:FSmallEta}
  F_{\s L}(\eta) &=& \GGG{L+\frac{D}{2}}{\s+\frac{D}{2}, 1-\s-\frac{D}{2}}
  \sum_{n=0}^\infty
  \GGG{\s+\frac{D}{2}+n, 1-\s-\frac{D}{2}+n}{1+n, L+\frac{D}{2}+n}
  \left(\frac{1-i\eta}{2}\right)^n ,
  \nn \\ & &
  \quad {\rm for\;} \left|\frac{1-i\eta}{2}\right| < 1 ,
}
while for $|(1-i\eta)/2| > 1$ it has the Laurent expansion:
\eqn{ \label{eq:FLargeEta}
  F_{\s L}(\eta) &=& \GGG{L+\frac{D}{2}, 2\s+D-1}{\s+\frac{D}{2}, L+\s+D-1}
  \GGG{2-D-2\s}{1-\s-\frac{D}{2}, 2-L-\s-D}
  \nn \\ & & \times
  \left(\frac{i \eta - 1}{2}\right)^{\s+(D-2)/2}
  \sum_{n=0}^\infty \GGG{1-\s-\frac{D}{2}+n, 2-L-\s-D+n}{1+n, 2-D-2\s+n}
  \left(\frac{2}{1-i\eta}\right)^n
  \nn \\ & & + \left( \s \to -(\s +D -1) \right),
  \quad {\rm for\;} \left|\frac{1- i \eta}{2}\right| > 1 .
}
The asymptotic expansion of $F_{\s L}(\eta)$ for large $|\eta|\gg 1$
with $\s$, $D$, and $L$ fixed is given by (\ref{eq:FLargeEta}); the
leading terms are
\eqn{
  F_{\s L}(\eta) &=&
  \GGG{L+\frac{D}{2}, 2\s+D-1}{\s+\frac{D}{2}, L+\s+D-1}
  \exp\left[i\frac{\pi}{2}\left(\s+\frac{D-2}{2}\right) \right]
  \left(\frac{\eta}{2}\right)^{\s+(D-2)/2}
  \nn \\ & & + \left( \s \to -(\s +D -1) \right) + \dots.
}
Sub-leading terms in the Laurant expansion (those with $n > 0$ in
(\ref{eq:FLargeEta})) are negligible when $|\eta| \gg 1$ and
$|\eta| \gg (L-\s)$. In this limit
\eqn{
  (1 + \eta^2)^{-(D-2)/4} &=& \eta^{-(D-2)/2} + O(\eta^{-1}) ,
  \nn \\
  \left[\frac{1-i\eta}{1+i\eta}\right]^{(L+(D-2)/2)/2}
  &=& \exp\left[i\frac{\pi}{2}\left(L+\frac{D-2}{2}\right)\right] ,
}
and so the full mode functions have the asymptotic expansion
\eqn{
  u_{\s L}(\eta) &=& \frac{N_{\s L}}{2^{\s+(D-2)/2}}
  \GGG{L+\frac{D}{2}, 2\s+D-1}{L+\s+D-1, \s+\frac{D}{2}}
  \exp\left[i\frac{\pi}{2}(L+\s+D-2)\right] (\eta)^{\s}
  \left[1 + O\left(\frac{(L-\s)}{\eta}\right)\right]
  \nn \\ & &
  + (\s \to - (\s+D-1)) ,
  \quad {\rm for\;} |\eta| \gg 1, \;\; |\eta| \gg L-\s .
}

\section{WKB approximation of Klein-Gordon modes}
\label{app:WKB}

In this appendix we derive the WKB approximations to the Klein-Gordon
modes (\ref{eq:WKB}). The d'Alembertian in the coordinates (\ref{eq:metric})
is
\eq{
  \ell^2 \Box = -(1+\eta^2) \d_\eta^2 - D \eta \d_\eta
  + \frac{1}{(1+\eta^2)} \nabla^2_{S^{D-1}} ,
}
where $\nabla^2_{S^{D-1}}$ is the scalar Laplacian on $S^{D-1}$ with
unit radius. Expanding the mode functions as
$\phi_{\s \vL}(x) = \ell^{(2-D)/2} u_{\s L}(\eta) Y_\vL(\vx)$ gives
\eq{ \label{eq:foru}
  (1+\eta^2) u_{\s L}''(\eta)
  + D \eta  u_{\s L}'(\eta)
  + f(\eta) u_{\s L}(\eta) =  0 ,
}
where primes denote derivatives with respect to $\eta$ and
\eq{ \label{eq:anotherf}
  f(\eta) = \frac{L(L+D-2)}{(1+\eta^2)} + M^2\ell^2 .
}
We now insert the ansatz
\eq{ \label{eq:ansatz}
  u_{\s L}(\eta) = K e^{A(\eta)} ,
}
and anticipate that $A(\eta) = A_0(\eta) + A_1(\eta) + \dots$, with
successive terms suppressed by a large parameter. Inserting (\ref{eq:ansatz})
into (\ref{eq:foru}) we obtain
\eq{
  (1+\eta^2)\left[A''(\eta) + (A'(\eta))^2\right]
  + D \eta A'(\eta) + f(\eta) = 0 .
}
At lowest order in the WKB approximation we keep only the term
$(A'_0(\eta))^2$; this yields
\eq{
  A'_0(\eta) = \pm i \left[\frac{ f(\eta) }{(1+\eta^2)} \right]^{1/2} .
}
We see that the large parameter in this WKB expansion is
$\sqrt{f(\eta)}$, and that we have just solved for the $O(f(\eta))$ part
of (\ref{eq:foru}). To obtain the sub-leading term $A_1(\eta)$ we
solve the $O(\sqrt{f(\eta)})$ part of (\ref{eq:foru}):
\eq{
  (1+\eta^2)\left[A''_0(\eta) + 2 A'_0(\eta) A'_1(\eta) \right]
  + D \eta A'_0(\eta) = 0,
}
from which we obtain
\eq{
  A'_1(\eta) = - \half \left[ \frac{D \eta}{1+\eta^2}
    + \frac{A''_0(\eta)}{A'_0(\eta)} \right]
  = \frac{d}{d\eta} \log\left[(1+\eta)^{-D/4}(A'_0(\eta))^{-1/2}\right] .
}
We note that
\begin{equation}
\Bigl| \frac{A'_1(\eta)}{A_0'(\eta)} \Bigr| \le \frac{A_{10}(\eta)}{\sqrt{f(\eta)}},
\end{equation}
where\footnote{Here we used $|F + \frac{f'}{2f}| \le |F| + |\frac{f'}{2f}| \le |F| + \frac{|f'|}{2L(L-D+2)} (1 + \eta^2)$ for $F(\eta) =  \frac{D\eta}{1+\eta^2}$.},
\begin{equation}
A_{10}(\eta)  = \frac{D}{2}  \frac{\eta}{\sqrt{1 + \eta^2}} 
\end{equation}
is independent of $L,M$.  This (and similar results at higher order) shows 
that our WKB expansion is uniformly valid at large $f(\eta)$.

Collecting our results we have
\eq{ \label{eq:uanswer}
  u_{\s L}(\eta) \approx \frac{1}{\sqrt{2}} (1+\eta^2)^{(1-D)/4}
  \left[ f(\eta) \right]^{-1/4} e^{\pm i \Upsilon(\eta) },
}
with $\Upsilon(\eta)$ the anti-derivative of
$\left[ f(\eta) / (1+\eta^2) \right]^{1/2}$.
The normalization is easily obtained by computing
the Klein-Gordon norm (\ref{eq:KGnorm}) of (\ref{eq:uanswer}).  The higher order terms give corrections to (\ref{eq:uanswer}) that are much smaller than (\ref{eq:uanswer}) in the limit of large $f$.

\section{Linearization formulae}
\label{app:linearization}

In this appendix we construct linearization formulae for the Bunch-Davies
2-point functions; these formulae follow immediately from the linearization
formula of the Euclidean Green's function.
To construct the linearization formula for Euclidean Green's
function we must recall some facts about the Euclidean Green's
function $\D_\s(x,y) = \D_\s(Z)$, where $Z := Z(x,y)$ is the $SO(D+1)$-invariant
distance between $x$ and $y$. This information is presented in more
detail in \cite{Marolf:2010nz}. Recall that $\D_\s(Z)$ may be written
as a Mellin-Barnes integral
\eq{ \label{eq:Delta}
  \D_{\s}(Z) = \ell^{2-D} \int_\nu \psi_\s(\nu)
  \Gamma(-\nu) \left(\frac{1-Z}{2}\right)^\nu ,
}
with
\eq{
  \psi_\s(\nu) := \frac{1}{(4\pi)^{\a+1/2}}
  \GGG{-\s+\nu, \s+2\a+\nu, \half-\a-\nu}{\half+\a+\s, \half-\a-\s} .
}
Here $\int_\nu\dots$ denotes a contour integral in the complex $\nu$ plane
traversed from $-i\infty$ to $+i\infty$ within the strip $\s < \Re\nu < 0$.
The measure $d\nu/(2\pi i)$ is assumed.
Because the sphere radius $\ell$ enters only as a multiplicative
constant we will set $\ell=1$ for now and restore it at the end of
the appendix.
It is convenient to use $\a := (D-1)/2$ to keep track of the spacetime
dimension. In Pauli-Villars regularization we subtract from the
unregulated Green's function a linear combination of Green's functions
organized such that all UV divergences cancel. We may write the regulated
Green's function as
\eqn{
  \D^{\rm reg}_\s(Z) = \D_\s(Z) + \sum_{i=1}^{[D/2]} C_i \D_{\rho_i}(Z)
  =
  \int_\nu \psi^{\rm reg}_\s(\nu)
  \Gamma(-\nu) \left(\frac{1-Z}{2}\right)^\nu ,
}
with
\eq{ \label{eq:psireg}
  \psi^{\rm reg}_\s(\nu) :=
  \psi_\s(\nu) + \sum_{i=1}^{[D/2]} C_i \psi_{\rho_i}(\nu) .
}
Here the $\rho_i$ are mass parameters corresponding to the PV masses
$M_i^2$. The values of the coefficients $C_i$ are not unique, though their
values are constrained such that the desired cancellations occur, and
they are bounded in the limit $M_i^2\to\infty$ \cite{bogoliubov:1980aa}.
For finite PV masses the function $\psi^{\rm reg}_\s(\nu)$ is analytic in
$\nu$ in the strip ${\rm Re\,}\s < {\rm Re}\,\nu < \half$ in all dimensions.
At large $M^2\gg 1$ the function $\psi_\s(\nu)$ has the asymptotic
behavior
\eq{ \label{eq:psilimit}
  \psi_\s(\nu) = \frac{M^{2\a-1+2\nu}}{(4\pi)^{\a+1/2}}\G{\half-\a-\nu}
  \left(1+O(M^{-2})\right) ;
}
it follows that in the limit $M_i^2\to\infty$ the regulated Green's
function reduces to the unregulated Green's function at finite separations.
Further details may be found in \S3.1-2 of \cite{Marolf:2010nz}.

The product of two regulated Euclidean Green's functions is given
by
\eq{ \label{eq:product}
  \D_{\s_1}^{\rm reg}(Z)\D_{\s_2}^{\rm reg}(Z)
  = \int_{\nu_1}\int_{\nu_1}
  \psi_{\s_1}^{\rm reg}(\nu_1)\psi_{\s_2}^{\rm reg}(\nu_2)\,
  \GG{-\nu_1, -\nu_2} \left(\frac{1-Z}{2}\right)^{\nu_1+\nu_2} .
}
One can easily invert (\ref{eq:Delta}) to find
\eq{ \label{eq:invert}
  \left(\frac{1-Z}{2}\right)^{\nu}
  = 2 (4\pi)^{\a+1/2} \int_\mu
  \GGG{\half+\a+\nu, \mu-\nu}{1+2\a+\mu+\nu, -\nu}(\mu+\a) \D_{\mu}(Z) .
}
Combining (\ref{eq:product}) and (\ref{eq:invert}) we immediately
obtain the expression
\eq{ \label{eq:DeltaLin}
  \D_{\s_1}^{\rm reg}(Z)\D_{\s_2}^{\rm reg}(Z) =
  \int_\mu f(\mu) \D_\mu(Z)
}
with $f(\mu)$ given by
\eq{ \label{eq:fmu}
  f(\mu) := 2 (4\pi)^{\a+1/2}(\mu+\a)
  \int_{\nu_1}\int_{\nu_2}
  \psi_{\s_1}^{\rm reg}(\nu_1)\psi_{\s_2}^{\rm reg}(\nu_2)
  \GGG{-\nu_1, -\nu_2, \half+\a+\nu_1+\nu_2, \mu-\nu_1-\nu_2}
  {-\nu_1-\nu_2, 1+2\a+\mu+\nu_1+\nu_2} .
}
In the complex $\mu$ plane the function $f(\mu)$ decays exponentially
away from the real axis, and is analytic in the strip
$\Re(\s_1+\s_2) < \Re \mu$. The contour of integration in
(\ref{eq:DeltaLin}) 
lies within the strip $\Re(\s_1+\s_2) < \Re \mu< 0$.

Next we perform some simple manipulations, the utility of which will
become clear later. Let us denote the Pauli-Villars parameters associated
with $\D_{\s_1}^{\rm reg}(Z)$ by $C_{1i}$, $M_{1i}^2$, and $\rho_{1i}$ respectively,
and likewise those parameters associated with $\D_{\s_2}^{\rm reg}(Z)$
by $C_{2j}$, $M_{2j}^2$, and $\rho_{2j}$. Let us further define
\eq{
  f_{\beta_1\beta_2}(\mu) :=
  2 (4\pi)^{\a+1/2}(\mu+\a)
  \int_{\nu_1}\int_{\nu_2}
  \psi_{\beta_1}(\nu_1)\psi_{\beta_2}(\nu_2)
  \GGG{-\nu_1, -\nu_2, \half+\a+\nu_1+\nu_2, \mu-\nu_1-\nu_2}
  {-\nu_1-\nu_2, 1+2\a+\mu+\nu_1+\nu_2}
}
(this is just (\ref{eq:fmu}) with unregulated $\psi_\s(\nu)$ functions).
By expanding the $\psi_{\s_i}^{\rm reg}(\nu_i)$ as in (\ref{eq:psireg})
we may write
\eq{
  f(\mu) = f_{\s_1\s_2}(\mu)
  + \sum_{i=1}^{[D/2]} C_{1i} f_{\rho_{1i}\s_2}(\mu)
  + \sum_{j=1}^{[D/2]} C_{2j} f_{\s_1\rho_{2i}}(\mu)
  + \sum_{i=1}^{[D/2]}\sum_{j=1}^{[D/2]}
  C_{1i} C_{2j} f_{\rho_{1i}\rho_{2j}}(\mu) .
}
Consider the terms $f_{\rho_{1i}\rho_{2j}}(\mu)$ in the last sum of this equation.
In these terms
we would like to move the contours of integration into the region
$\Re \nu_1 < \half-\a$, $\Re \nu_2 < \half -\a$. For $D\ge 4$ we
can only do so at the cost of encountering poles. Performing this
manipulation yields
\eq{
  f_{\rho_{1i}\rho_{2i}}(\mu) = h_{\rho_{1i}\rho_{2j}}(\mu)
  + \sum_{n=0}^{[(D-4)/2]} c^n_{\rho_{1i}\rho_{2j}}
  \GGG{\half+\a+\mu+n}{\half+\a+n, \half+\a+\mu-n} 2(\mu+\a) .
}
Here $h_{\rho_{1i}\rho_{2j}}(\mu)$ is just $f_{\rho_{1i}\rho_{2j}}(\mu)$
with the contours satisfying $\Re \nu_1 < \half-\a$, $\Re \nu_2 < \half -\a$
as desired, and the sum is due to the residues of the poles encountered
for $D \ge 4$. The $c^n_{\rho_{1i}\rho_{2j}}$ are coefficients that do not
depend upon $\mu$:
\eq{
  c^n_{\rho_{1i}\rho_{2j}} :=
  (4\pi)^{\a+1/2}
  \int_\nu \psi_{\rho_{1i}}(\nu)\psi_{\rho_{2j}}\left(-\half-\a-\nu-n\right)
  \GG{-\nu,\half+\a+\nu+n} .
}
Defining yet another quantity,
\eq{
  f^{\rm van}(\mu) :=
  \sum_{i=1}^{[D/2]} C_{1i} f_{\rho_{1i}\s_2}(\mu)
  + \sum_{j=1}^{[D/2]} C_{2j} f_{\s_1\rho_{2i}}(\mu)
  + \sum_{i=1}^{[D/2]}\sum_{j=1}^{[D/2]}
  C_{1i} C_{2j} h_{\rho_{1i}\rho_{2j}}(\mu) ,
}
we may write $f(\mu)$ as
\eqn{ \label{eq:fagain}
  f(\mu) &=& f_{\s_1\s_2}(\mu) + f^{\rm van}(\mu)
  \nn \\ & & +
  \sum_{n=0}^{[(D-4)/2]} \left[
      \sum_{i=1}^{[D/2]}\sum_{j=1}^{[D/2]}
      C_{1i} C_{2j} c^n_{\rho_{1i}\rho_{2j}} \right]
    \GGG{\half+\a+\mu+n}{\half+\a+n, \half+\a+\mu-n} 2(\mu+\a) .
    \nn \\
}
None of these manipulations have altered any of the salient
features of $f(\mu)$.

The purpose of these manipulations has been to isolate the part
of $f(\mu)$ which diverges in the limit where the Pauli-Villars
regulator masses $M_{ij}^2\to\infty$. In this limit:
\begin{enumerate}
  \item the function $f_{\s_1\s_2}(\mu)$, which is independent of
    the regulator masses, survives unaltered,
  \item every term in $f^{\rm van}(\mu)$ decays like a negative power
    of at least one regulator mass, and so it vanishes in the limit,
  \item the terms in the sum on the last line of (\ref{eq:fagain})
    diverge like positive power of a regulator mass.
\end{enumerate}
In particular, for the spacetime dimensions of interest for our
computation we may write these divergences explicitly:
\eqn{
  f(\mu) &=& f_{\s_1\s_2}(\mu) + f^{\rm van}(\mu),
  \quad\quad\quad\quad\quad\quad\quad
  \quad\quad\quad\quad\quad\quad\quad
  \quad\quad\quad\quad\;
  (D=2,3) \nn \\
  f(\mu) &=& f_{\s_1\s_2}(\mu) + f^{\rm van}(\mu) + c_0 2 (\mu+\a) ,
  \quad\quad\quad\quad\quad\quad\quad
  \quad\quad\quad\quad\quad\;\;
  (D = 4,5) \nn \\
  f(\mu) &=& f_{\s_1\s_2}(\mu) + f^{\rm van}(\mu)
  + c_0 2 (\mu+\a) + c_1 2 (\mu+\a)[-\mu(\mu+2\a)]  ,
  \quad (D = 6) .
}
Here $c_0$ and $c_1$ are coefficients that diverge with a Pauli-Villars
mass like $c_0 \sim \log M$ in $D=4$, $c_0 \sim M$ in $D=5$, and
$c_0 \sim M^2\log M $ and $c_1 \sim \log M$ in $D=6$. We have not
yet taken the $M_{ij}^2\to\infty$ limit, but we have organized $f(\mu)$
so that this limit will be quite easy.

We are almost done. Noting the relations
\eqn{
  \int_\mu 2(\mu+\a) \D_\mu(Z) &=& \frac{\delta(Z-1)}{(1-Z^2)^{\a-1/2}} ,
  \\
  \int_\mu 2(\mu+\a)[-\mu(\mu+2\a)] \D_\mu(Z)
  &=& \Box \left[\frac{\delta(Z-1)}{(1-Z^2)^{\a-1/2}}\right] ,
}
and restoring the radius $\ell$,
we may finally record the linearization formulas for the dimensions
of interest:
\eqn{
  \D^{\rm reg}_{\s_1}(Z)\D^{\rm reg}_{\s_2}(Z)
  &=& \ell^{2-D} \int_\mu (f_{\s_1\s_2}(\mu) + f^{\rm van}(\mu)) \D_\mu(Z) ,
  \quad\quad\quad\quad\quad\quad\quad\quad\quad\quad
  \quad\quad\;\; (D=2,3) \nn \\
  \D^{\rm reg}_{\s_1}(Z)\D^{\rm reg}_{\s_2}(Z)
  &=& \ell^{2-D} \int_\mu (f_{\s_1\s_2}(\mu) + f^{\rm van}(\mu)) \D_\mu(Z)
  + \ell^{2D-4}c_0 \frac{\delta(Z-1)}{(1-Z^2)^{\a-1/2}} ,
  \quad\quad
  (D=4,5) \nn \\
  \D^{\rm reg}_{\s_1}(Z)\D^{\rm reg}_{\s_2}(Z)
  &=& \ell^{2-D} \int_\mu (f_{\s_1\s_2}(\mu) + f^{\rm van}(\mu)) \D_\mu(Z)
  \nn \\ & &
  + \ell^{2D-4} c_0 \frac{\delta(Z-1)}{(1-Z^2)^{\a-1/2}}
  + \ell^{2D-2} c_1 \Box \left[\frac{\delta(Z-1)}{(1-Z^2)^{\a-1/2}}\right] ,
  \quad (D=6) .
}

The linearization formulae for Lorentzian Green's functions may
be found by the usual analytic continuation. For the time-ordered
Green's functions we analytically continue $Z \to Z+i\epsilon$.
The resulting linearization formulae are given in (\ref{eq:Glin}).
To obtain the linearization formulae for the Wightman Green's
function we analytically continue with the cut prescription
$Z \to Z + s(x_1,x_2) i \epsilon$, where
$s(x_1,x_2) = +(-)$ if $x_1$ is in the future (past)
of $x_2$.  This cut prescription removes contact terms, so the result
is the simple expression (\ref{eq:Wlin}).


\addcontentsline{toc}{section}{Bibliography}
\bibliographystyle{JHEP}
\bibliography{./bibliography}

\end{document}